\documentclass[acmsmall, screen]{acmart}
\definecolor{Lightgray}{gray}{0.9}
\usepackage[font=footnotesize, labelfont=bf]{caption}

\AtBeginDocument{%
  }

\setcopyright{none}
\copyrightyear{0}
\acmYear{2026}
\acmDOI{XXXXXXX.XXXXXXX}
\acmJournal{TOMS}
\acmISBN{XXXX}

\begin{document}
\title{NektarIR: A Domain-Specific Compiler for High-Order Finite Element Operations on Heterogeneous Hardware}
\author{Edward Erasmie-Jones}
\email{edward.erasmie-jones@kcl.ac.uk}
\orcid{0009-0003-1006-6900}
\affiliation{%
    \institution{King's College London}
    \city{London}
    \country{United Kingdom}
}
\author{Giacomo Castiglioni}
\authornote{Work done while the author was employed at Swiss National Supercomputing Centre (CSCS), ETH Zurich}
\email{giacomo.castiglioni@huawei.com}
\orcid{0000-0003-0278-6951}
\affiliation{%
    \institution{Computing Systems Lab, Huawei Zurich Research
Center}
    \city{Zurich}
    \country{Switzerland}
}

\author{David Moxey}
\email{david.moxey@kcl.ac.uk}
\orcid{0000-0003-2186-8397}
\affiliation{
    \institution{King's College London}
    \city{London}
    \country{United Kingdom}
}

\begin{abstract}
   Modern high performance computing (HPC) applications must target heterogeneous hardware. This requires significant work to ensure domain specific implementations translate to highly performant kernels across a range hardware types and vendors, each requiring bespoke optimization to make use of the specific target architecture. Through the development of a domain specific compiler built with the multi-level intermediate representations (MLIR) project, one can express a high-level, close to the specific domain, abstraction that is progressively lowered to a low, close to metal, abstraction. At each intermediate representation (IR), appropriate optimizations can be applied without costly analysis due to the knowledge embedded in the domain specific IRs. We apply this method to the construction of discrete differential operators for use in spectral/\textit{hp} element method solvers for computational fluid dynamics (CFD). Here, the performance is driven by a small set of common finite element operators that are composed to create kernels for the discrete differential operators used to solve weak partial differential equations. We create our own MLIR dialect to represent these operators and implement a bespoke lowering pipeline to facilitate the just-in-time compilation of these kernels for both CPU and GPU architecture and illustrate performance comparisons with the Nektar++ spectral/\textit{hp} element framework.
    
\end{abstract}




\maketitle
\section{Introduction}

The development of robust, efficient solvers based on high-order and spectral/\textit{hp} element methods has been an area of sustained and growing interest over recent decades. The adoption of higher-order polynomial expansions within elements brings well-understood numerical advantages: substantially reduced levels of dispersion and dissipation compared to low-order schemes, making these methods particularly attractive for applications such as computational fluid dynamics where the accurate advection of fine-scale energetic structures is critical~\cite{sherwinbook, reviewSIAM}. At the same time, and perhaps more acutely in the context of the present hardware landscape, high-order methods offer compelling computational advantages. Although the cost per degree of freedom grows with polynomial order, the resulting finite element operators are characterised by dense, compact kernels with high arithmetic intensity — a property of considerable value when the primary bottleneck in modern processor architectures is memory bandwidth rather than raw floating-point throughput~\cite{roofline}.

This interplay between numerical efficiency and hardware utilisation has motivated a substantial body of work targeting three closely related ideas: matrix-free formulations of finite element operators, which avoid the explicit construction of assembled global or local elemental matrices; the exploitation of tensor-product structure through sum-factorisation~\cite{Orszag1980SpectralGeometries}, which reduces operator evaluation from $\mathcal{O}(P^{2d})$ to $\mathcal{O}(P^{d+1})$ operations; and the explicit exploitation of vector instruction sets to approach the peak floating-point throughput of modern CPUs and GPUs. These ideas in combination have been demonstrated to yield impressive computational performance in frameworks including deal.II~\cite{2025:dealII}, Neko~\cite{Jansson2024Neko:Dynamics}, NekRS~\cite{2022:NekRS}, Dune~\cite{Bastian2021TheDevelopments}, MFEM~\cite{Andrej2024High-performanceMFEM} and CEED~\cite{Kolev2021EfficientMethods}.

A significant limiting factor in much of this prior work, however, is a reliance on meshes composed exclusively of quadrilateral elements in two dimensions and hexahedral elements in three dimensions. Although such structured element types admit a natural tensor-product structure and thus the greatest potential for arithmetic optimisation, the generation of all-hexahedral meshes remains an open problem in computational geometry, particularly for complex three-dimensional domains of engineering and scientific relevance. The difficulty is compounded at high order, where curved elements that conform to complex boundary geometries are required, introducing additional challenges in mesh generation and element validity~\cite{Kirilov2024High-OrderMeshes}. For the challenging geometries anticipated at exascale — including turbomachinery, cardiovascular flows, and high-Reynolds-number external aerodynamics — unstructured meshes combining tetrahedra, triangular prisms, and pyramids are not merely convenient, but necessary. Although matrix-free evaluation of finite element operators on these simplicial and hybrid element types is achievable and has been the study of our recent work~\cite{Eichstadt2023EfficientDimensions, Moxey2020EfficientElements}, it is an additional parameter among an already complex range of performance-critical aspects: the performance of a given finite element kernel is a sensitive function of a large number of parameters: polynomial order $P$, element type, quadrature rule, and the nature of the geometric mapping (affine vs. deformed), as well as the specifics of the underlying hardware including vector width, cache hierarchy, and the availability of fused multiply-add instructions or tensor cores.

There is, therefore, a fundamental challenge of performance portability across diverse and evolving hardware targets, which remains largely unaddressed. This portability challenge is not merely logistical.  Optimal implementations for these combinations are in practice obtained through careful hand-tuning of kernels, as demonstrated in a number of prior studies~\cite{ Jansson2024Neko:Dynamics, Xing2026Architecture-awareMeshes, Andrej2024High-performanceMFEM}. The result is a fundamental tension between performance and generality: hand-tuned kernels attain near-peak performance but are hardware-specific and require re-optimisation as architectures evolve, while portable implementations sacrifice the performance gains that high-order methods are intended to deliver. This tension is particularly acute in the context of adaptive simulations, where the mesh, polynomial order, or both may change dynamically over the course of a computation. In $hp$-adaptive methods~\cite{hpadaptivepaper2,hp-adaptivepaper1}, elements may be individually enriched or coarsened in response to \emph{a posteriori} error estimates, leading to configurations in which the optimal operator evaluation strategy — including the appropriate degree of loop unrolling, vectorisation width, and quadrature configuration — varies not just between runs, but within a single simulation. Static, ahead-of-time compilation of operator kernels is fundamentally ill-suited to this setting: one cannot anticipate all combinations of polynomial order and element type at compile time without generating a combinatorially large library of pre-compiled kernels. This motivates the use of just-in-time (JIT) compilation, in which operator kernels are generated, optimised, and compiled at runtime in response to the specific configuration encountered. JIT approaches have begun to appear in the finite element literature — notably in Firedrake~\cite{Rathgeber2016Firedrake:Abstractions} and FEniCS~\cite{Baratta2025DOLFINx:Environment}, where custom form compilers generate optimised low-level code from high-level problem specifications. However, these approaches have primarily targeted portability and correctness, with performance optimisation remaining a secondary concern, and also rely on a source-to-source translation strategy which requires the maintenance and development of a complex form compiler hierarchy in order to target the desired hardware platforms.

Multi-Level Intermediate Representation (MLIR)~\cite{Lattner2020MLIR:Law}, an extensible compiler infrastructure developed within the LLVM ecosystem~\cite{Lattner2004LLVM:Transformation}, offers a promising but as yet largely unexplored route to address this challenge for scientific computing. MLIR was introduced to provide reusable infrastructure for the creation of intermediate representations (IR) of software implementations at abstractions levels higher (or lower) than LLVM IR and has since been adopted as the compiler backbone for a number of prominent AI and deep learning frameworks such as IREE~\cite{Liu2022TinyIREE:Deployment}, StableHLO~\cite{stablehlo} and torch-mlir~\cite{LLVM_Torch-MLIR}. Its design around a hierarchy of extensible dialects — each capturing a different level of abstraction, from high-level linear algebra operations through to hardware-specific vector and GPU instructions — makes it a natural fit for the kind of structured, parametric computation that characterises finite element operator evaluation. MLIR facilitates the translation of IR written using MLIR dialects to LLVM IR, which can be efficiently compiled to multiple hardware types and vendors, including via JIT compilation strategies, thereby ensuring extensibility of the application.

While MLIR has been used in the development of domain-specific compilers for stencil based computations~\cite{Gysi2021Domain-SpecificSimulation}, its application to other scientific computing domains, and to high-order finite element methods in particular, has received little attention in the literature to date. The vast majority of MLIR-related publications address machine learning model compilation or hardware accelerator targeting for tensor operations arising in neural network inference. The structural similarities between these tensor operations and the sum-factorised kernels of the spectral/$hp$ element method — both involving sequences of contractions over structured, parametric index spaces — suggest that MLIR's optimisation infrastructure may translate favourably to the finite element setting. However, this connection has not been systematically explored, and it is not clear \emph{a priori} whether the abstractions and optimisation passes developed in the ML context will yield the arithmetic intensity and memory access patterns required for high-performance finite element operator evaluation on general, unstructured meshes.

In this paper, we address this gap directly. We present \emph{NektarIR}: a new domain-specific compiler for the generation of highly efficient high-order finite element kernels for the use in high-order spectral/\textit{hp} element simulations for computational fluid dynamics.  NektarIR aims to facilitate the generation and just-in-time (JIT) compilation of kernels for common finite element operations used in the construction of various discrete differential operators for both CPU and GPU targets. Our approach defines a high-level representation of finite element operator evaluation in terms of a custom MLIR dialect, which is progressively lowered through a sequence of transformations — including vectorisation, and hardware-specific code generation — to produce kernels optimised for a given polynomial order and element type at runtime. We demonstrate that this approach can match and in some cases exceed the performance of hand-tuned, statically-compiled kernels across a range of element types and polynomial orders, while retaining the flexibility to adapt seamlessly to changing simulation configurations. The framework is developed and evaluated within the Nektar++ spectral/hp element library~\cite{Moxey2020Nektar++:,Cantwell2015Nektar++:Framework}, and the kernels are benchmarked against the state-of-the-art implementations described in~\cite{Moxey2020EfficientElements} and~\cite{Xing2026Architecture-awareMeshes}.

The content presented in this report describes considerations for efficient implementations of high-order finite element operations on heterogeneous hardware and the design of a domain-specific MLIR dialect. Several of the design considerations we encountered for our dialect, and the implementation our IR transformations at different abstraction levels are considerations that may be encountered in the design of any new MLIR dialect. As such, we believe the work herein is of interest for both scientific computing and compiler design communities.  

The remainder of the paper is structured as follows. The finite element operations our dialect represents are discussed in section \ref{sec:background} and the design and content of our dialect and the transformations we have added is discussed further in section~\ref{NektarIR}. In section~\ref{results} we present an investigation into the performance of our compiler and overhead cost of the compilation pipeline. For runtime performance comparison, we compare the throughput of the kernels generated by NektarIR to those available in Nektar++ on both CPU and GPU targets. Finally, we conclude with a brief summary of our work in~\ref{conclusion}.

\section{Background}
\label{sec:background}
We begin with a brief background to the present work. Firstly, we highlight where the finite element operations we represent as MLIR dialect operations arise in the context of the spectral/\textit{hp} element method applied to the Helmholtz equation: an elliptic PDE that underpins the CFD solvers implemented in codes such as Nektar++. We then briefly discuss how MLIR and LLVM simplify the development of domain-specific compilers and some of the components of existing MLIR dialects.

\subsection{The spectral/\textit{hp} element method}\label{section:sem}
The starting point for the derivation of the finite element operations that form the foundation of the abstraction in our domain-specific compiler is the finite element discretizations of the Helmholtz equation. Efficient implementations of the Helmholtz operator and Helmholtz solvers are critical: in many high-order CFD solvers, including Nektar++, we leverage a velocity correction scheme \cite{Karniadakis1991High-orderEquations} in which operator splitting is used to decouple velocity and pressure, and which involves the solution of a Poisson equation and $d$ Helmholtz equations, where $d=2,3$ is the spatial dimension. The Helmholtz equation on a two- or three-dimensional domain $\Omega$ is given by
\begin{equation}
    \nabla^{2}u -\lambda u = f(x)
    \label{eqn: helmholtz}
\end{equation}
where $u:\Omega \to \mathbb{R}$ and $f:\Omega \to \mathbb{R}$ are functions for the desired solution and forcing term, respectively, and $\lambda > 0$. Following a standard finite element approach, the solution domain $\Omega$ is subdivided into a mesh of non-overlapping elements. As we illustrate in the upcoming section, these elements may be of different shapes, and this changes the nature of the elemental operations. In this work, we will restrict ourselves consider only 3-dimensional regions that are either hexahedra or tetrahedra. Within each elemental region, the solution $u$ is discretised and given by
\begin{equation}
    u^{\delta}(\mathbf{x}) = \sum_{i=0}^{P}\hat{u}_{i}\phi^{e}_{i}(\mathbf{x}),
    \label{eq: discreteu}
\end{equation}
where each $\hat{u}_{i}$ are the unknown expansion modes to be solved for and $\phi_{i}^{e}(\mathbf{x})$ are the basis polynomials and $P$ is the order of the expansion. To obtain the weak form of the Helmholtz equation, Equation~\eqref{eqn: helmholtz} is multiplied by a test function, $v$, and integrated over the solution domain to obtain
\begin{equation}
    (\nabla v, \nabla u)_{\Omega^{e}} - \lambda(v,u)_{\Omega^{e}} = -(v,f)_{\Omega^{e}},
    \label{eq:weak_helm}
\end{equation}
where, for simplicity, we have assumed homogeneous Neumann boundary conditions to eliminate any boundary terms. $\Omega^{e}$ is a single elemental region in the mesh and the inner product $(\nabla v, \nabla u)_{\Omega^{e}}$ is given by
\begin{equation}
    (u,v)_{\Omega^{e}} = \int_{\Omega^{e}}u(\mathbf{x})v(\mathbf{x}) d\mathbf{x}.
    \label{innerprodcts}
\end{equation}
By substituting the discretisation of $u$ in ~\eqref{eq: discreteu},  applying a Galerkin approach to use the same discretisation for $v$, and restricting the set of points where the discretizations are evaluated to a set of quadrature points within each element, Equation~\eqref{eq:weak_helm} can be rewritten in matrix form as 
\begin{equation}
    \mathbf{L}^{e}\hat{\mathbf{u}}^e + \lambda \mathbf{M}^{e}\hat{\mathbf{u}}^{e} = -(\mathbf{B}^{e})^{T}\mathbf{W}^{e}\mathbf{f}^{e}
    \label{eqn:matrix_helm}
\end{equation}
where $\mathbf{L}^e$ is the discrete form of the weak Laplacian, namely $(\nabla v, \nabla u)_{\Omega^{e}}$, and $\mathbf{M}^{e}$ is the discrete form of $(v,u)_{\Omega^{e}}$, often referred to as the mass matrix. The matrices $\mathbf{B}^{e}$ and $\mathbf{W}^{e}$ are the basis and weights matrices respectively and correspond to the matrix form of two of the elemental operators which are discussed in more detail below. The vectors $\hat{\mathbf{u}}^{e}$ and $\mathbf{f}^{e}$ respectively contain the expansion modes from the discretization in Equation~\eqref{eq: discreteu} and the forcing function from Equation~\eqref{eqn: helmholtz} evaluated at the quadrature points. The discrete, weak Helmholtz operator $\mathbf{H}^{e}$ can be obtained by factorizing $\hat{\mathbf{u}}$ on the left-hand side of Equation~\eqref{eqn:matrix_helm} and is given by
\begin{equation}
    \mathbf{H}^{e} = (\mathbf{L}^{e}\ + \lambda \mathbf{M}^{e}).
    \label{eqn:weakmatrixhelm}
\end{equation}
Both $\mathbf{L}^e$ and $\mathbf{M}^{e}$ can be further decomposed and given by
\begin{equation}
    \mathbf{L}^{e} = (\mathbf{D}^{e}_{x_{i}}\mathbf{B}^{e})^T\mathbf{W}^{e}\mathbf{D}^{e}_{x_{i}}\mathbf{B}^{e} \quad \text{and} \quad \mathbf{M}^{e} = (\mathbf{B}^{e})^{T}\mathbf{W}^{e}\mathbf{B}^{e}
    \label{laplacian}
\end{equation}
where the repeated index $i$ is used to indicate summation over the different coordinate directions. 

The size of the basis matrix $\mathbf{B}^{e}$ is given by the product of the number of quadrature points, $N_{q}$ and the number of expansion modes, $M$. Its entries are given by
\begin{equation}
    [\mathbf{B}^{e}]_{ij} = \phi^{e}_{j}(\boldsymbol{\xi}_{i}),
    \label{basis matrix}
\end{equation}
namely the basis polynomials $\phi^{e}_{j}$ evaluated at the quadrature points, $\boldsymbol{\xi}_{i}$. The structure of $\mathbf{B}^{e}$ and the total number of expansion modes depends on the type of basis polynomials used for the local approximation and the shape of the elemental region, as discussed further in Section~\ref{sec: elementalops}. 
The weights matrix $\mathbf{W}^{e}$ is a diagonal matrix where the non-zero entries are given given by the product of the quadrature weights and the determinant of the Jacobian transformation from the standard element to an element in the mesh.

The differentiation matrix $\mathbf{D}^{e}_{x_{i}}$ corresponds to collocation differentiation in an elemental region $\Omega^{e}$ and, using the chain rule, can be given in terms of the derivatives in the standard element $\Omega_{st}$ as 
\begin{equation}
    \mathbf{D}^{e}_{x_{i}} = \mathbf{\Lambda^{e}}\left(\frac{\partial \xi_{j}}{\partial x_{i}}\right)\mathbf{D}^{e}_{\xi_{j}}
    \label{eqn:physderiv}
\end{equation}
where the repeated index $j$ is used to indicate summation over the coordinate directions in the standard element. The elemental matrix $\mathbf{\Lambda^{e}}\left(\frac{\partial \xi_{j}}{\partial x_{i}}\right)$ is a diagonal matrix containing $\left(\frac{\partial \xi_{j}}{\partial x_{i}}\right)$ evaluated at the quadrature points and $\mathbf{D}^{e}_{\xi_{j}}$ is the differentiation matrix for the standard element.

To solve the weak Helmholtz equation over the entire domain, the local contributions to the solution, $\mathbf{\hat{u}}^{e}$, must be assembled into a global system which can be solved as a system of equations. This assembly operation is not discussed as a part of this work. 

\subsection{Elemental Operations}\label{sec: elementalops}
In this section we will elaborate on the presentation of the elemental operations found in the discrete form of the weak Helmholtz equation shown in equation~\eqref{eqn:weakmatrixhelm}. In particular, we will highlight the difference between the elemental operators for two elemental region shapes, namely the hexahedron and tetrahedron. Following the presentation given in Chapter 3 of~\cite{sherwinbook}, the standard hexahedron is defined as
\begin{equation}
    \Omega_{\text{hex}} = \{(\xi_{1},\xi_{2},\xi_{3}) \mid \xi_{1},\xi_{2},\xi_{3} \in [-1, 1]^{3} \},
    \label{stdhex}
\end{equation}
where the constant bounds on each coordinate $\xi_{i}$ are convenient when integrating over the domain, as required when evaluating the inner products shown in equation~\eqref{eq:weak_helm}. On the other hand, the standard tetrahedron is defined as
\begin{equation}
    \Omega_{\text{Tet}} = \{(\xi_{1},\xi_{2},\xi_{3}) \mid -1\leq\xi_{1},\xi_{2},\xi_{3}, \xi_{1}+\xi_{2}+\xi_{3} \leq -1 \},
    \label{stdtet}
\end{equation}
where the bounds on the coordinates are seen to be non-constant, leading to inconvenient bounds on the integrals over the domain. Through repeated use of the Duffy transformation, the tetrahedron can be expressed in terms of a local, collapsed coordinate system with constant bounds and hence given by 
\begin{equation}
    \Omega_{\text{Tet}} = \{(\eta_{1},\eta_{2},\eta_{3}) \mid \eta_{1},\eta_{2},\eta_{3} \in [-1, 1]^{3} \},
    \label{collapsedtet}
\end{equation}
where the $\eta$ coordinates are given by
\begin{equation}
    \eta_{1} = \frac{2(1+\xi_{1})}{-\xi_{2}-\xi_{3}}, \quad \eta_{2} = \frac{2(1+\xi_{2})}{1-\xi_{3}}-1, \quad \eta_{3} = \xi_{3}.
    \label{eta-space}
\end{equation}
While integration over the tetrahedron defined using the $\eta$ coordinates yields more convenient integration bounds, the definition of $\eta_{1}$ and $\eta_{2}$ introduces singularities for some values of $\xi_{2}$ and $\xi_{3}$ that must be accounted for. In general, these are handled by using a set of quadrature points for $\xi_{1}$, $\xi_{2}$, $\xi_{3}$ that do not include the endpoints of the interval $[-1,1]$ such as Gauss-Radau points. Nonetheless, the Jacobian of the mapping to the collapsed coordinate system requires special consideration for both numerical integration and differentiation in the standard tetrahedron. The collapsed coordinates also changes the structure of the basis matrix, $\mathbf{B}$. The first elemental operation, namely the the \textbf{backward transform} performs a projection from the modal (expansion) space to the physical space. In matrix form, it corresponds to the action of the basis matrix and is given by
\begin{equation}
    \mathbf{u} = \mathbf{B}\hat{\mathbf{u}},
\end{equation}
where $\mathbf{u}$ is a vector containing the approximation of $u$ from equation~\eqref{eqn: helmholtz} evaluated at the quadrature points within an elemental region. The basis matrix contains the basis polynomials evaluated at each quadrature point. In summation form, the backward transform is given by
\begin{equation}
    u_{n} = \sum_{p = 0}^{M}\hat{u}_{p}\phi_{p}(\boldsymbol{\xi_{n}}) \quad \forall n
    \label{eq:bwdstd}
\end{equation}
where $\phi_{p}(\boldsymbol{\xi_{n}})$ are the basis polynomials and $\boldsymbol{\xi_{n}}=(\xi_{1i}, \xi_{2j},\xi_{3k})$ are the $n = i\times j \times k $ quadrature points. While it is possible to consider the backward transform as a simple matrix-vector product, a common choice of basis in the spectral/\textit{hp} element method is one which can be expressed in terms of a tensor product of 1-dimensional basis tensors. In this case, the total order $M$ is a function of the expansion order in each coordinate direction $P_{1}$,$P_{2}$ and $P_{3}$. For the remainder of this work we will only consider the case where the basis admits a tensor product expansion. In this case, we utilize the ``sum-factorisation" technique~\cite{Orszag1980SpectralGeometries} optimization technique. For a hexahedral region, the total number of modes, $M(P_{1},P_{2},P_{3}) = P_{1}P_{2}P_{3} $ and equation~\eqref{eq:bwdstd} can be rewritten as
\begin{equation}
    u_{ijk} = \sum_{p = 0}^{P_{1}-1}\psi^{a}_{p}(\xi_{1i})\left(\sum_{q = 0}^{P_{2}-1}\psi^{a}_{q}(\xi_{2j})\left(\sum_{r = 0}^{P_{3}-1}\hat{u}_{pqr}\psi^{a}_{r}(\xi_{3k})\right)\right) \quad \forall i,j,k,
    \label{bwdsumfachex}
\end{equation}
where $\psi^{a}_{p}(\xi_{1i})$, $\psi^{a}_{q}(\xi_{2j})$ and $\psi^{a}_{r}(\xi_{3k})$ are the 1-dimensional basis tensors of the tensor product expansion, evaluated at the quadrature points in the $i$-th, $j$-th and $k$-th directions respectively. The ``sum-factorisation" form of the backward transform for a tetrahedral region is given by 
\begin{equation}
    u_{ijk} = \sum_{p = 0}^{P_{1}-1}\psi^{a}_{p}(\eta_{1i})\left(\sum_{q = 0}^{P_{2}-1 - p}\psi^{b}_{pq}(\eta_{2j})\left(\sum_{r = 0}^{P_{3}-1 - p - q}\hat{u}_{m(p,q,r)}\psi^{c}_{pqr}(\eta_{3k})\right)\right) \quad \forall i,j,k,
    \label{bwdsumfactet}
\end{equation}
where the quadrature points given by $\eta_{1i}$, $\eta_{2j}$ and $\eta_{3k}$ are the collapsed $\eta$ space shown in equation~\eqref{eta-space} and $P_{1} \leq P_{2} \leq P_{3}$. The specific forms of $\psi^{a}_{p}$, $\psi^{b}_{pq}$ and $\psi^{c}_{pqr}$ depend on the nature of the dynamics being simulated and the mesh of the solution domain. These are further discussed in~\cite{sherwinbook} and~\cite{Moxey2020EfficientElements}.
The mapping $m(p,q,r)$ in the subscript of the expansion mode $\hat{u}$ is given by
\begin{equation}
    m(p',q',r') = \sum_{p=0}^{p'}p\sum_{q=0}^{q'-p}q\sum_{r=0}^{r'-p-q}r.
    \label{tetmap}
\end{equation}
The expanded form of equation~\eqref{tetmap} is readily obtained by expanding each summation for an arbitrary $p'$, $q'$ and $r'$. The next elemental operation is the \textbf{inner product}, which evaluates the discrete form of $(u,v)_{\Omega^{e}}$, as shown in equation~\eqref{innerprodcts}. In matrix form, it is given by
\begin{equation}
    \mathbf{\hat{u}} = \mathbf{B}^{T}\mathbf{W}\mathbf{u},
\end{equation}
where $\mathbf{W}$ is a diagonal matrix containing the quadrature weights required for numerical integration and the Jacobian determinants from the coordinate transformation from the physical to reference element. $\mathbf{B}^T$ is the transpose of the basis matrix, and the inner product over the hexahedron can be expressed using ``sum-factorisation" as
\begin{equation}
    \hat{u}_{pqr} = \sum_{i = 0}^{Q_{1}-1}w_{1p}\psi^{a}_{p}(\xi_{1i})\left(\sum_{j = 0}^{Q_{2}-1}w_{2q}\psi^{a}_{q}(\xi_{2j})\left(\sum_{k = 0}^{Q_{3}-1}\lvert J \rvert{u}_{ijk}w_{3k}\psi^{a}_{r}(\xi_{3k})\right)\right) \quad \forall p,q,r,
    \label{eqn: iprodhex}
\end{equation}
where $Q_{1}$, $Q_{2}$ and ${Q_{3}}$ are the number of quadrature points in each coordinate direction respectively. $w_{1i}$, $w_{2j}$ and $w_{3k}$ are the quadrature weights evaluated at the $i$-, $j$- and $k$-th quadrature points respectively. For curvelinear or \textit{deformed} elements, the Jacobian determinant will vary at each quadrature point and the form given above is restricted to straight-sided elements where $\lvert J \rvert$ is constant. In a tetrahedral region, the ``sum-factorisation" form of the inner product of the approximation and the test function is given by
\begin{equation}
    \hat{u}_{m(p,q,r)} = \sum_{i = 0}^{Q_{1}-1}w_{1p}\psi^{a}_{p}(\eta_{1i})\left(\sum_{j = 0}^{Q_{2}-1}\tilde{w}_{2q}\psi^{b}_{pq}(\eta_{2j})\left(\sum_{k = 0}^{Q_{3}-1}\lvert J \rvert{u}_{ijk}\tilde{w}_{3k}\psi^{c}_{pqr}(\eta_{3k})\right)\right) \quad \forall p,q,r,
    \label{eqn: iprodtet}
\end{equation}
where the mapping $m(p,q,r)$ is once again required in the subscript of $\hat{u}$ on the left-hand side. The composition of the backward transform with the inner product on a vector of expansion coefficients $\mathbf{\hat{u}}$ corresponds to the mass matrix, $\mathbf{M}$, from equation~\eqref{eqn:weakmatrixhelm}. For both the backward transform and the inner product, temporary storage is required to hold the result of any of the intermediary summations. The matrices $\mathbf{D}_{x_{i}}$ and $\mathbf{D}_{\xi_{i}}$ represent the action of differentiation in the physical and standard element, respectively. For instance, the partial derivative of $u$ in the $\xi_{1}$ direction within the standard element, or the \textbf{standard derivative}, is given by
\begin{equation}
    \frac{\partial u}{\partial{\xi}_{1}}(\xi_{1i}, \xi_{2j}, \xi_{3k}) = \mathbf{D_{\xi_{1}}}\mathbf{u} = \sum_{r=0}^{Q_{1}-1}\left.\frac{dh_{r}(\xi_{1})}{d\xi_{1}}\right\rvert_{\xi_{1i}}\left(\sum_{s=0}^{Q_{2}-1}h_{s}(\xi_{2j})\left(\sum_{t=0}^{Q_{3}-1} h_{t}(\xi_{3k})u_{rst}\right)\right) \quad \forall i,j,k,
    \label{eqn:stdderiv}
\end{equation}
where $u_{rst}$ is the $(rst)$-th component of the vector containing the approximation of $u$ evaluated at the quadrature points, and $h_{r}$, $h_{s}$ and $h_{t}$ represent Lagrange polynomials of order r, s and t respectively. The derivative of the basis polynomials is similarly given by $\mathbf{D}_{\xi_{i}}\mathbf{B}$ and is required as part of the discrete Laplacian in equation~\eqref{eqn:weakmatrixhelm}. To evaluate the derivative of the approximation $u$ in the physical element, i.e the action of $\mathbf{D}_{x_{i}}$, the diagonal matrix $\mathbf{\Lambda}(\partial \xi_{j}/\partial x_{i})$ containing $\partial \xi_{j}/\partial x_{i}$ evaluated at each quadrature point is required due to the mapping from the standard to physical element. We also use the diagonal matrix $\mathbf{\Lambda}$ to express the derivative in terms of the local coordinates in the tetrahedron, with for instance
\begin{equation}
    \mathbf{D}_{\xi_{i}} = \mathbf{\Lambda}\left(\frac{\partial \eta_{j}}{\partial \xi_{i}} \right)\mathbf{D}_{\eta_{j}}
\end{equation}
where the index $j$ is repeated on the right-hand side to indicate summation over the collapsed coordinate directions within the tetrahedron. 

The total amount of computational work required for each of the above operations is not only related to the number of modes and quadrature points, but also to the shape of the elemental region, as highlighted by the different loop bounds on both the backward transform and inner product in equations ~\eqref{bwdsumfactet} and ~\eqref{eqn: iprodtet}. This is also the case for the derivative, where the $\eta$-to-$\xi$ mapping adds additional work in the evaluation of the standard and physical derivative. All this information is available at compile-time and can be used to inform compiler optimizations such as loop-unrolling, loop-fusion and memory pooling for any temporary allocations required by the inner product and backward transform. All the operations presented above act on a single element but may also be batched over a collection, or block, of alike elements, each of which having the same shape, number of quadrature points and expansion order.

\subsection{LLVM and MLIR} \label{LLVMandMLIR}

Before moving onto our presentation of NektarIR, in this section we first expand briefly on compiler technologies and their use of intermediate representations (IR).

A traditional compiler pipeline is composed of three phases: a front end, optimizer and backend. During the front end, the source language is parsed and an abstract syntax tree (AST) is created to represent the input implementation, which is then converted to an IR that is passed to the optimizer and backend. The optimizer is responsible for rewriting and transforming the IR to improve the runtime performance of the application. In traditional compilers, the transformations the optimizer can perform, such as dead and redundant code elimination, loop fusion, and vectorization, are limited by the low, assembly-like abstraction level of the IR and may require sophisticated analysis. This low-level representation may also be unable to express all domain-specific metadata necessary for performance-critical optimizations. The compiler backend facilitates the code-generation of hardware specific machine code by mapping the IR to the hardware's instruction set. A low-level IR is advantageous here as it provides a simpler mapping to the hardware instructions and allows the compiler backend to facilitate hardware specific optimizations, such as register allocation and instruction scheduling~\cite{LLVMBook}. The IR at the optimization phase is often independent of both the source language and hardware target and the extensibility of the compiler is only limited by which languages it has a front end for and which hardware targets it can map the IR to. 

LLVM~\cite{Lattner2004LLVM:Transformation} is an open-sourced collection of compiler development tools which simplifies the development of new and extensible compiler infrastructure with its own IR, LLVM IR, that is independent of the input source language and can be compiled to a wide range of hardware targets. Additionally, the framework has a just-in-time (JIT) compilation driver, the LLVM Execution Engine. LLVM and LLVM IR are used in a range of modern compilers and languages, such as the C and C++ compiler Clang~\cite{ClangDevelopersClang:LLVM}, the Julia programming language~\cite{Bezanson2017Julia:Computing} and Rust~\cite{klabnik2026rust}. LLVM IR is a low-level assembly-like language with a large semantic gap compared to high-level languages. To bridge this gap, a recent compiler framework developed under the LLVM umbrella, namely the multi-level intermediate representation (MLIR) project~\cite{Lattner2020MLIR:Law}, was designed to enable the creation of intermediate representations of higher (or any) abstraction level. In MLIR, this is achieved through the use of \textit{dialects} that represent different programming constructs at various levels of abstraction. This adds a multi-level IR rewriting stage to the optimization phase of the compiler pipeline, where the IR is converted between dialects using MLIR conversion passes that progressively lower the abstraction level of the IR. At any stage, MLIR passes can also be used for optimizing transformations of the IR. MLIR provides a low-level LLVM dialect, which can be translated to LLVM IR before backend code-generation using LLVM.

MLIR allows for the expression of high abstraction level representations of scientific computing applications, such as those in domain-specific languages (DSLs), as their own  MLIR dialects, or with preexisting high-level dialects that are already provided. An MLIR dialect consists of a set of operations, types and attributes. A dialect operation is a unit of code in MLIR, akin to an instruction in a classical IR. The IR created using MLIR dialects must obey static single-assignment (SSA) form, meaning each value defined by an operation can only be defined once. The type of a value contains compile-time information which is used to verify the correctness of the IR, and within the MLIR passes which transform the operation. Custom dialect types can also be used to inject domain-specific information into the IR. Dialect attributes are used to attach compile-information to operations. While the number of operations, types and attributes available in MLIR is not fixed and can be conveniently expanded upon by developers for their own application, already available operations can be readily reused in IR created for any application and operations defined as part of new MLIR dialects can freely make use of existing types and attributes. Common dialects in MLIR-based compilers include the \textit{func} dialect for function abstractions, the \textit{scf} dialect for abstractions of structured control flow, the \textit{memref} dialect for operations on shaped regions of memory, and the \textit{arith} dialect for arithmetic operations. Other, more specialized dialects include the \textit{vector} and \textit{gpu} dialects for vendor-independent abstractions of SIMD and GPU programming models, respectively. There are also dialects designed for specific types of optimizations, such as the \textit{affine} dialect for polyhedral loop optimization and the \textit{linalg} dialect for tiled linear algebra computations.

\begin{figure}[t!]
    \centering
    \includegraphics[width=0.9\linewidth]{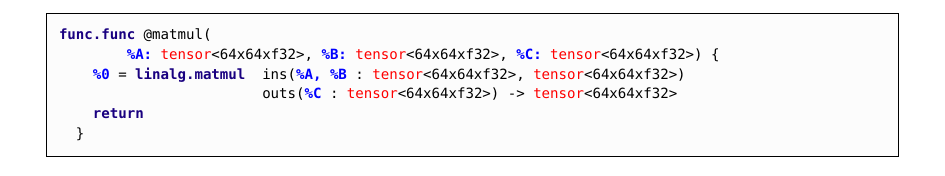}
    \caption{IR example of a matrix-matrix product kernel using the Linalg and func dialects}
    \label{fig:LinalgExample}
    \Description{A code-listing showing an example of a matrix multiplication kernel using the linalg and func dialects in MLIR.}
\end{figure}

Figure~\ref{fig:LinalgExample} contains an example of the IR corresponding to a basic matrix-matrix product written using the \textit{func} and \textit{linalg} dialects. Here, the arguments of the \texttt{func.func} operation each have a tensor type, which is a built-in type in MLIR. Values with tensor types are immutable, and operations that act on and produce tensors obey value semantics. Through a process known as \textit{bufferization}, tensor types are converted into memrefs, which represent shaped regions of memory. Operations on values with memref types obey reference semantics. Furthermore, the \texttt{linalg.matmul} operation uses a ``destination-passing style'' and takes in the destination tensor as an ``outs'' operand. This means that although the operation produces a new tensor to represent the result, a copy of the destination tensor \texttt{\%C} may not be materialized in the IR, as long as there are no other conflicting uses of the tensor, thereby avoiding additional allocations. After bufferization, and an IR transformation that converts the \texttt{linalg.matmul} operation to a set loops, the resulting IR corresponds to the familiar triple-loop, pointer-based matrix-multiplication implementation.  
\section{NektarIR} \label{NektarIR}
In this section we will present our MLIR dialect, NektarIR (abbreviated as {\tt nir}), and how we use dialect types and attributes to create a high-level abstraction of the elemental operators described in the previous section. We also present the compiler pipeline from a C++ based front end to JIT compiled elemental operator kernels via MLIR and LLVM for both CPU and GPU hardware targets. Here, we will present new MLIR passes that utilize domain-specific information to perform optimizing transformations on the IR at various abstraction levels and transform the IR from a single, high-level abstraction to hardware specific kernels for both CPU and GPU architectures.

\subsection{The NektarIR Dialect}
At the NektarIR dialect level, we have aimed to create an abstraction that closely resembles the mathematical representation of the elemental operators in the spectral/\textit{hp} element method. The operations in our dialect correspond to the elemental operators, while the types describe the local expansion and the shape of the elemental region. We use attributes to attach domain-specific information to operations from other MLIR dialects that we use during our IR transformations after the representation in {\tt nir}. 

\subsubsection{Types}
All elemental operations act on and produce values with a custom \texttt{block} type. This type is an unbufferized type (i.e. it has value semantincs). The aim of this type is to represent a grouping of elements within a mesh that share a common shape type, polynomial order, quadrature rule and spatially regular vs. constant Jacobian. Values of the \texttt{block} type represent either the data corresponding to the expansion coefficients, or the approximation evaluated at the quadrature points in physical space; i.e. $\hat{\mathbf{u}}$ and $\mathbf{u}$ respectively in eq.~\eqref{eq:bwdstd}. To distinguish between the two states, the \texttt{block} type encodes either the quadrature rule or basis type, which may optionally be defined in each coordinate direction to denote tensor-product decompositions.

\begin{figure}
    \centering
    \includegraphics[width=\linewidth]{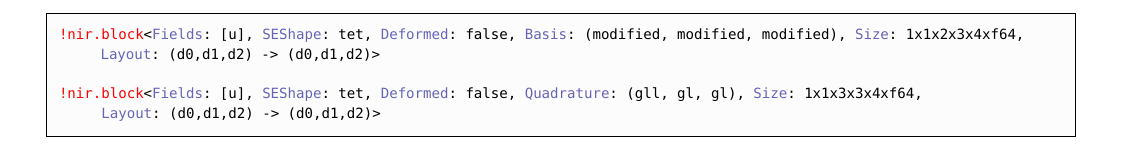}
    \caption{Examples of the \texttt{block} type}
    \label{fig:blocktype}
    \Description{A code listing showing the type definitions of the NektarIR block type in the IR}
\end{figure}

Figure~\ref{fig:blocktype} contains examples of the \texttt{block} type as it would appear in the IR when describing a grouping of alike elements:

\begin{itemize}
  \item \texttt{SEShape} defines the shape of the element represented within this \texttt{block}: in this case, tetrahedral elements.
  \item \texttt{Basis} is used to specify that this \texttt{block} contains data corresponding to the expansion coefficients. This is defined as a tuple of strings which highlights the tensor-product decomposition of this basis.
  \item \texttt{Quadrature} is used to specify that this \texttt{block} contains data corresponding to the approximation evaluated at the quadrature points. This is defined as a string which describes the quadrature rule in each coordinate direction. For instance, the specifier \texttt{gll} represents the Gauss-Lobatto-Legendre quadrature.
  \item \texttt{Deformed} is a boolean flag which when \texttt{true} denotes the Jacobian mapping is spatially varying for each element, and when \texttt{false} is constant.
  \item \texttt{Fields} defines a string list of variable names in order to support multi-component data. In this case we represent a single $u$-component of velocity.
  \item The \texttt{Size} attribute encodes information required to compute the storage size of the block: namely, the number of components, number of elements, the order of the expansion in each direction, namely $P_{1}$, $P_{2}$ and $P_{3}$, and the data type. For a \texttt{block} with a quadrature rule specifier, the final dimensions specified by the size will instead correspond to the number of quadrature points in each direction.
  \item Finally the \texttt{Layout} attribute encodes the data ordering within the \texttt{block}, which we elaborate on below.
\end{itemize}

The layout of the data is an important parameter to encode as this allows for specialised implementations that can exploit specific ordering of data, particularly via SIMD instructions as shown in e.g.~\cite{Moxey2020EfficientElements}. This is encoded through the \texttt{Layout} attribute, which accepts a 3-dimensional permutation map. The parameters denote the number of fields, number of elements and total number of expansion coefficients (or quadrature points), which can be used to represent the ordering of the data in the underlying memory region. For instance, in Figure~\ref{fig:blocktype} the underlying data is contained in a 3-dimensional buffer of the native MLIR \texttt{memref} type. In this case, the mapping is the identity, but these may appear in a different order depending on the source of the input data and the use-case. This reordering is referred to as \textit{interleaving}, and will be discussed further in the upcoming sections.

Note that the overall storage size for the block is not necessarily computed by simple multiplication of all of the dimensions specified by the \texttt{Size} attribute. For non-tensorial elements such as a tetrahedron, the final dimension in the data buffer corresponds to the result of the mapping $M(P_{1}, P_{2}, P_{3})$ in Equation~\eqref{tetmap}, which importantly is always less than $P_{1}P_{2}P_{3}$. In this sense, the element shape encoded in the \texttt{block} type acts as an abstraction for the sparsity of the underlying data. The element shape, its deformity, the dimensions specified in the size of the block and the layout are all used to inform the transformation of the elemental operations into IR from other MLIR dialects.

\subsubsection{Operations}
The operations in our MLIR dialect fall into three categories: operations representing the elemental operations in the spectral/\textit{hp} element method as described in Section \ref{sec: elementalops}, operations for the creation and manipulation of values with \texttt{block} type and operations used to facilitate the conversion, or bufferization, of values from a \texttt{block} type to a \texttt{memref} type. We begin with the elemental operations:

\begin{itemize}
    \item \texttt{bwd} represents the backward transform, as in Equation~\eqref{eq:bwdstd}.
    \item \texttt{inner} represents inner product of the approximation and the test function, as in Equation~\eqref{innerprodcts}.
    \item \texttt{standard\_deriv} represents the derivative with respect to the local coordinates in the reference element, as in Equation~\eqref{eqn:stdderiv}.
    \item \texttt{phys\_deriv} represents the derivative with respect to the physical coordinates in the mesh element, as in Equation~\eqref{eqn:physderiv}.
    \item \texttt{apply\_jw} represents the action of the weights matrix.
    \item \texttt{test} represents the action of transpose of the basis matrix, namely the basis of the expansion of the test function. This is the adjoint of the backward transform. 
    \item \texttt{test\_grad} represents the action of the derivative of the transpose of the basis matrix, namely the derivative of the basis of the test function             expansion.
    \item \texttt{add} represents elementwise addition of two \texttt{block} type values.
    \item \texttt{helmholtz} represents the Helmholtz operator, as in Equation~\eqref{eqn:weakmatrixhelm}.
    \item \texttt{mass} represents the mass matrix operator. 
\end{itemize}
In each case, the operations take in a \texttt{block} type and must produce values of \texttt{block} type as results. The dimension corresponding to the number of elements in a \texttt{block} can be considered as the ``batch" dimension and if the input block only contains a single element, the operation is prefixed by \texttt{elmnt}. Importantly, the elemental operations are designed to not use a destination-passing style, meaning that none of the elemental operations take in the expected destination for the result as an operand. Not using a destination-passing style allows us to express the composition of several of elemental operations, such as the Helmholtz operator, with a simple chain of our dialect operations. This removes the lifespan and aliasing of destination \texttt{blocks} as a consideration for IR transformations that act on our elemental operations, and leaves further memory considerations to a lower abstraction level. The other operands of the elemental operations, such as the basis data, quadrature weights or derivative matrices are expected to be values with the builtin MLIR \texttt{tensor} type.

The operations for the creation and manipulation of values with \texttt{block} type are modeled after operations with the same name in the MLIR \texttt{tensor} dialect, which contains operations for the creation and manipulation of values with \texttt{tensor} type. In our dialect, these include:
\begin{itemize}
    \item \texttt{empty\_block} an empty \texttt{block}.
    \item \texttt{insert\_slice} inserts a source \texttt{block} into a destination \texttt{block} at a specific offset and returns a copy of the destination \texttt{block} with the inserted slice.
    \item \texttt{extract\_slice} extracts a slice of a source \texttt{block} at a specific offset and returns the extracted slice.
\end{itemize}

Lastly, the operations used to facilitate the conversion, or bufferization, of values from a \texttt{block} type to a \texttt{memref} type are:
\begin{itemize}
    \item \texttt{block\_from\_memref} creates a value with a \texttt{block} type from a memory region and additional expansion data. After bufferization, the result \texttt{block} is replaced by the associated memref. 
    \item \texttt{materialize\_in\_destination} associates a value with a \texttt{block} type with a destination buffer. After bufferization, the input \texttt{block} is replaced by the destination buffer. 
\end{itemize}
These operations are modeled after the operations in the MLIR \texttt{bufferization} dialect which are used for the conversion from tensor to memref types. 

The operations we define as part of our MLIR dialect combine to create the highest level abstraction of an elemental operator kernel in our compiler pipeline. In the next section, we see how they are transformed as the IR gets converted to upstream MLIR dialects. 

\subsection{NektarIR Transformations}
\begin{figure}
    \centering
    \includegraphics[width=0.9\linewidth]{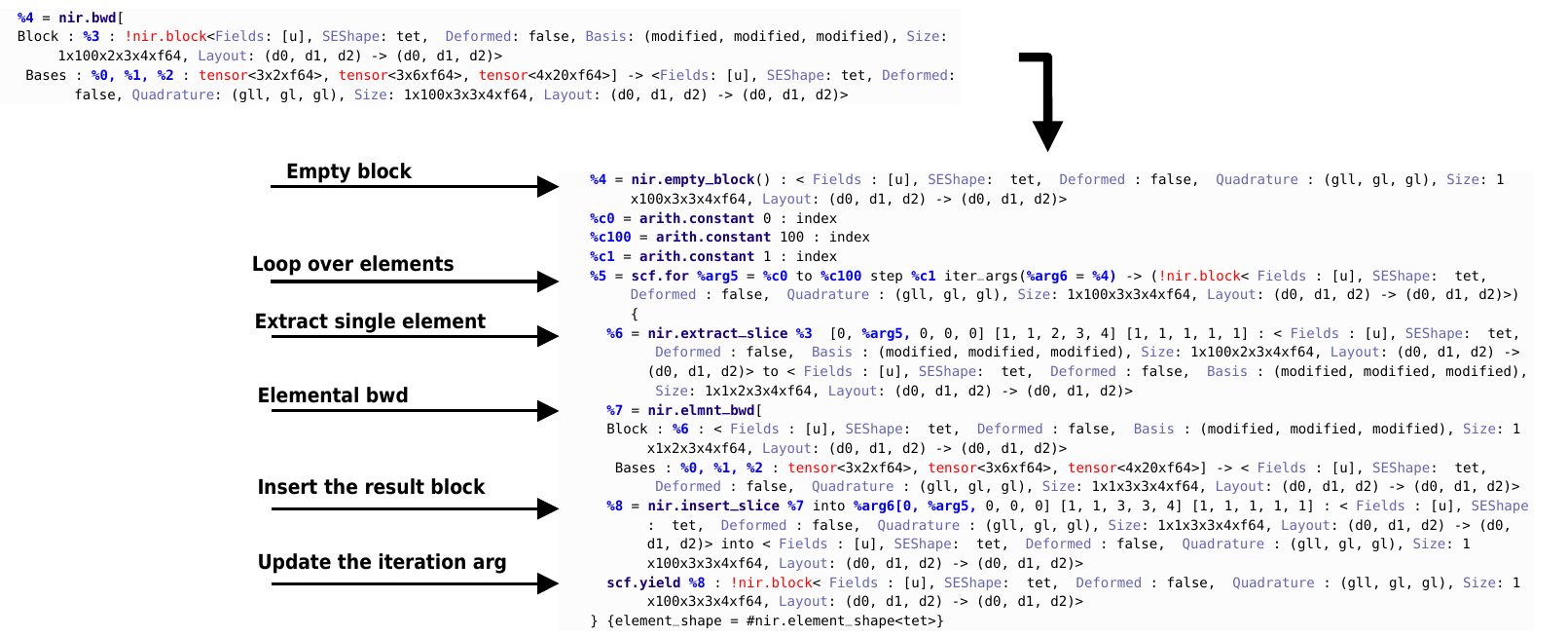}
    \caption{Conversion from block backward transform, \texttt{nir.bwd} to a loop over elements and an elemental backward transform, \texttt{nir.elemental\_bwd}.}
    \Description[Schematic of the IR transformation from a block backward transform to the elemental operation backward transformation]{Arrows indicate where new operations are created, including: a empty block operation, the new loop over elements, the extract block operation, the elemental backward transform operation, an insert slice operation and a materialization into the result buffer.}
    \label{fig:convertblocktoelemental}
\end{figure}

The IR transformations implemented in NektarIR are used to lower the IR from our dialect to preexisting MLIR dialects, perform loop transformations depending on the hardware target and optimize temporary memory allocations. The first transformation converts a batched elemental operation, such as \texttt{bwd} into a loop over the elements in the block and a \texttt{elmnt.bwd} operation. This transformation is highlighted in Figure \ref{fig:convertblocktoelemental}.

The representation of a batched \texttt{bwd} operation over the elements in the \texttt{block} using a loop over the elements and a \texttt{elmnt.bwd} operation has the following structure, as shown in Figure \ref{fig:convertblocktoelemental}:
\begin{itemize}
    \item A \texttt{empty\_block} operation is created to hold the result for all elements.
    \item The empty \texttt{block} is passed to the loop over elements as an induction variable to be updated with each iteration. 
    \item Within the loop, a \texttt{extract\_slice} operation returns a \texttt{block} corresponding to a single element in the mesh from the input to the \texttt{bwd} operation being transformed.
    \item A \texttt{elmnt.bwd} operation is created, which returns the result of the backward transform applied to a single element.
    \item The result of \texttt{elmnt.bwd} is inserted into the induction variable with a \texttt{insert\_slice} operation that returns the updated \texttt{block}.
    \item The updated \texttt{block}, namely the result of the \texttt{insert\_slice}, is yielded back to the loop updating the induction variable with the \texttt{block} holding the result. 
    \item The \texttt{scf.for} operation creates a value which represents the empty \texttt{block} that has now been filled with the result of the backward transform for each element. This \texttt{block} is equivalent to the result of the batched \text{bwd} operation. 
\end{itemize}

For more complicated batched operations, such as \texttt{helmholtz} and \texttt{mass}, the \texttt{elmnt} operations contained in the loop over elements will correspond to the sequence of elemental operations required to construct the operator as given in Equations~\eqref{eqn:weakmatrixhelm} for Helmholtz and~\eqref{laplacian} for mass. This ensures the \texttt{helmholtz} and \texttt{mass} operations are automatically generated as fused elemental operators, without the need for explicit kernel fusion or inlining. 

\begin{figure}
    \centering
    \includegraphics[width=\linewidth]{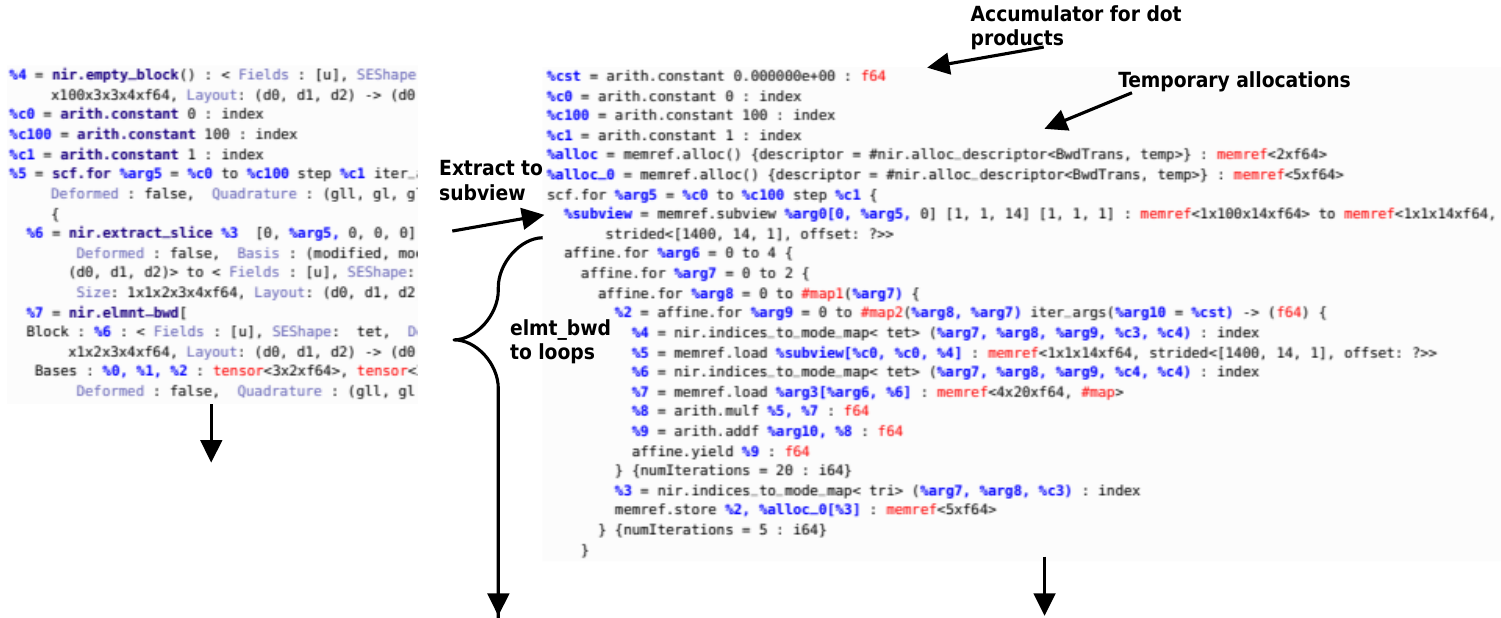}
    \caption{Transformation from NektarIR representation of the backward transform to the \texttt{affine} dialect and explicit loops. The loop structure resembles the expected implementation for the backward transform as described by Equation \eqref{bwdsumfactet}. The figure contains a section of the transformed IR, with arrows indicating that the IR continues.}
    \Description[IR transformation from NektarIR to upstream MLIR dialects]{An IR snippet showing the before and after of the IR corresponding to a backward transform operation in NektarIR as it is transformed to upstream MLIR dialects with explicit loops. Arrows indicate where operations corresponding to temporary allocations have been placed, as well as the result of the transformation on other NektarIR operations.}
    \label{fig:bwd as loops}
\end{figure}

After the lowering from batched to single element operations, the conversion of the \texttt{elmnt} operations and the bufferization of \texttt{block} types can take place. This lowering marks the start of the representation of the elemental operations using upstream MLIR dialects, rather than our dialect operations. The starting point for the lowering process is the conversion of the elemental operations to explicit loops, the bufferization of \texttt{block} and \texttt{tensor} types to memory buffers (i.e. \texttt{memref}s) and the conversion of operations manipulating \texttt{blocks} to the associated memref operation. This transformation is shown in Figure \ref{fig:bwd as loops} where the backward transform operation is converted to explicit loops represented using the \texttt{affine} dialect. At the explicit loop stage, the representation of the \texttt{bwd} operation is similar to what one would expect in a C++ implementation.

The main transformations of \texttt{nir} operations shown in both Figure \ref{fig:convertblocktoelemental} and Figure \ref{fig:bwd as loops} operations as part of the lowering to upstream MLIR dialects can be summarized by:
\begin{itemize}
    \item \texttt{block\_from\_memref} is replaced by the input buffer containing the coefficient data for all elements in the block. 
    \item \texttt{extract\_slice} is converted to a \texttt{memref.subview} which creates a subview of the input buffer which corresponds to coefficient data for a single element.
    \item \texttt{elmt\_bwd} is replaced by a collection of nested \texttt{affine.for} operations which represent for-loops. The structure of the loop nests correspond to the summation described by the sum-factorized backward transform operation shown in Equation~\eqref{bwdsumfactet}. Temporary allocations to hold the result of intermediate summations are generated as \texttt{memref.alloc} operations with \texttt{nir.alloc\_descriptor} attributes. These attributes are used for a buffer reuse transformation used to fuse temporary allocations for several elemental operations which do not overlap. 
    \item \texttt{insert\_slice} is replaced by the destination buffer as its purpose is only to update the result block, which now corresponds to the destination buffer.
    \item \texttt{materialize\_in\_destination} solely connects the result block with the pre-allocated result buffer and is removed. 
\end{itemize}

Once the elemental operator kernels are expressed using explicit loops from upstream MLIR dialects, we use a combination of both provided MLIR passes and our own to progressively lower our IR towards the hardware target, as discussed further in the upcoming section.

\subsection{Compiler Pipeline}
\begin{figure}
    \centering
    \includegraphics[width=\linewidth]{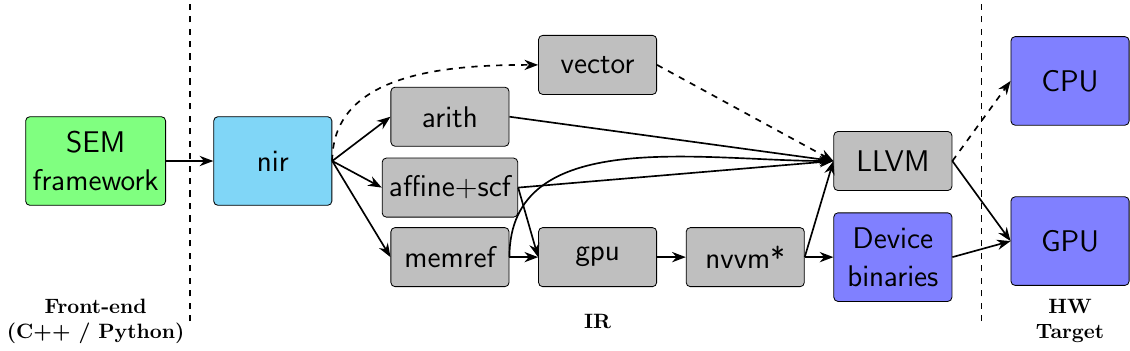}
    \caption{\textbf{The NektarIR lowering pipeline}. Schematic overview of the dialects visited by the IR for an elemental operation before compilation to a particular hardware target. The dashed arrows represent the additional step required for SIMD code-generation for CPU targets. (*) The nvvm dialect is a vendor specific dialect for NVIDIA GPUs; other vendor specific dialects exist and are used to lower to their devices. such as the rocdl dialect for AMD devices.}
    \Description[A schematic of an IR lowering pipeline]{The figure shows how IR created from a spectral/hp element framework gets converted as it is lowered towards the target hardware. Arrows connect nodes which correspond to individual MLIR dialects, starting from NektarIR (or nir for short) and ending at LLVM IR, visiting the vector and GPU dialects along the way. The final nodes correspond to the hardware targets, namely the CPU or the GPU.}
    \label{fig:loweringpipeline}
\end{figure}

NektarIR is designed independently of a front-end spectral/\textit{hp} element framework and, in the interest of re-usability, our dialect aims to represent the elemental operators in a manner that is close to the mathematics. An example of the end-to-end compiler pipeline for a spectral/\textit{hp} element (SEM) framework which uses NektarIR to generate its elemental operator kernels is shown in Figure \ref{fig:loweringpipeline}. Currently, we support the programmatic generation of elemental operators from C++ based front-ends and python support is in development. As part of the intermediate transformations in the lowering pipeline, we either utilize the \texttt{vector} dialect or the \texttt{gpu} dialect, depending on the hardware target.

\subsubsection{The Vector Dialect}
The MLIR \texttt{vector} dialect is a hardware-target independent abstraction of vector operations that can be used to target SIMD instruction sets. Values defined using vector operations have a \texttt{vector} type which is used as a virtual representation of actual hardware vectors, such as vector registers on a CPU. As highlighted in~\cite{Moxey2020EfficientElements}, it is natural to vectorize the elemental operations over the dimension corresponding to the number of elements, thereby allowing each elemental operation to act on a vector-width number of elements at once.  While there is already a vectorization pass available in MLIR as part of the \texttt{affine} dialect, any \texttt{nir} operation acting on a non-hyperrectangular shape (such as the tetrahedron), will define an iteration space that is incompatible with some \texttt{affine} operations, thereby making our IR incompatible with the preexisting vectorization pass. As such, the generation of vectorized IR is done during the transformation from block to elemental operations and elemental operations to loops. We represent this by extracting a block of vector-width number of elements to be passed to the elemental operation, and setting the step on the loop over elements to the vector-width as well. During the subsequent transformation to loops, the data layout component of the \texttt{block} type is critical, as it can be used to indicate that input data from a vector-width block of elements is stored contiguously in the underlying data buffer, meaning contiguous vector load and store operations can be used. For completeness, we also support vectorization without interleaved data, in which case the load and store operations that are generated as part of our lowering to loops are abstractions of strided load and store operations. Values defined by scalar loads from buffers represented as tensor typed operands in our elemental operations, such as the basis tensors or quadrature weights, are broadcast to the vector-width using \texttt{vector.broadcast} operations. Multiply-add patterns are fused to \texttt{math.fma} operations which are lowered to fused multiply-add intrinsics at the LLVM IR level.

\subsubsection{Targetting GPUs}\label{gpusection}
The GPU programming model is vastly different to the traditional CPU model and, to target GPU hardware, our IR needs to contain operations that correspond to specific aspects of GPU architecture. These include the grid, block and thread execution hierarchy, the global, shared and local memory structure as well as synchronous and asynchronous execution. Communication and synchronization between the host and the device is also necessary to generate launchable GPU kernels. In MLIR, these concepts are all contained in the vendor-independent \texttt{gpu} dialect. As the elemental operations acting on an element in a block are all independent of the other elements in the block, a natural parallelism in our kernels is over the elements. To express this in the IR, we use IR transformations to convert the loop over elements to a \texttt{gpu.launch} operation which takes in a grid configuration as well as buffers for allocation in either shared or private device memory. In our case, the specifics of the grid configuration, namely the grid and block sizes, depend on whether each thread is assigned to an element or to an expansion coefficient. The former case, which we refer to as ``threading through elements", is analogous to the SIMD approach described in the previous section in the sense that each thread in a block is assigned an element from the interleaved element block, and executes all the necessary instructions for that element. Temporary allocations, such as those that hold the result of intermediate elemental operations or the result of intermediate summations within an operation, must be expanded depending on the memory space they are assigned to. While it is possible to place these allocations in private or shared memory for individual elemental operations at low polynomial orders, this is not the case for composed operations such as the Helmholtz operator where the amount of temporary memory required exceeds available private and shared memory. These allocations are therefore expanded to the grid size and placed in global memory. The GPU kernels corresponding to several fused elemental operations, such as the mass matrix and Helmholtz operations, contain a large amount of instructions for each thread to carry out. This introduces a significant register pressure which may limit the possible parallelization of the kernel on the GPU device, thereby hindering performance. 

\begin{figure}
    \centering
    \includegraphics[width=\linewidth]{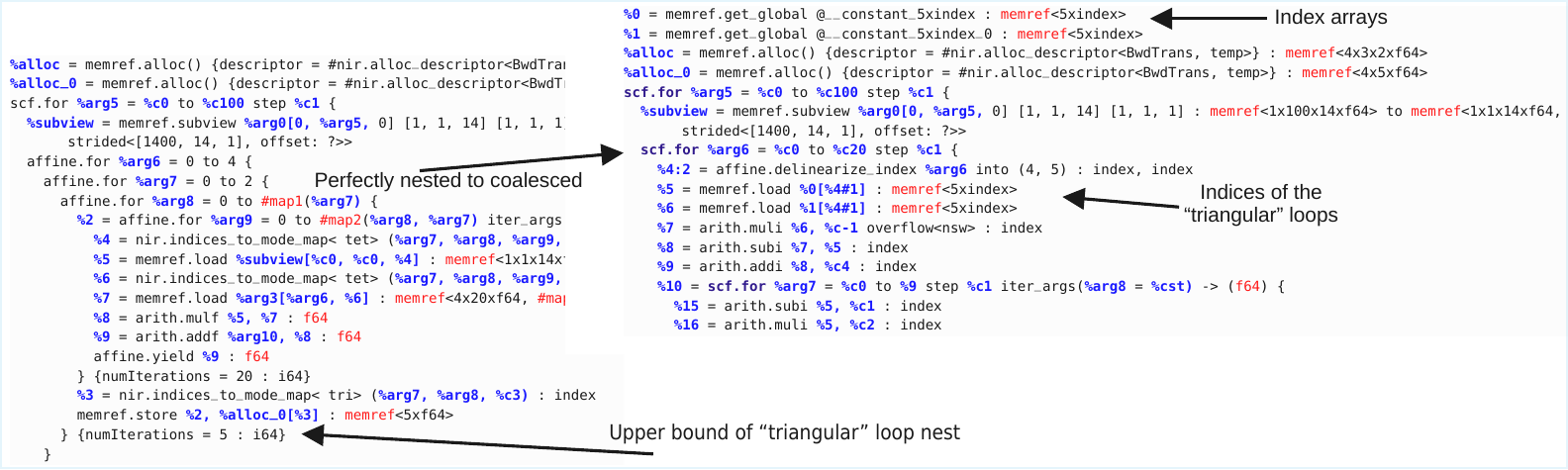}
    \caption{\textbf{Loop coalescing of a triangular loop nest using loop coalescing IR transformations}. The loops with non-constant upper bounds given by affine-maps are coalesced into a single loop. The upper bound of the resulting loop is obtained from an attribute attached to the innermost loop of the triangular loop nest (which is placed there when the loop is created as part of earlier transformations). Index arrays contain the indices for the coalesced loops.}
    \label{fig:loop coalescing}
    \Description[A IR snippet]{This IR snippet shows the before and after of a collection of for loops that are coalesced using loop coalescing IR transformations from NektarIR and upstream MLIR. Arrows indicate where arrays have been created as a look-up table for the recomputed indices as well as where they are loaded inside the new coalesced loop. Another arrow indicates where the total trip count, or number of iterations, of the triangular iteration space is stored for use in the transformation.}
\end{figure}

The alternative threading strategy, where each thread is mapped to an expansion coefficient, sets the grid size to the number of elements and the block size to the nearest multiple of the device warp size greater than the number of expansion coefficients. Here, the data is not interleaved and each expansion coefficient is stored contiguously. This threading strategy requires the IR corresponding to an elemental operation at the loop-level to change significantly, as each reduction must be computed using data accessed using indices determined by the thread index, rather than individual loops over the quadrature points or expansion order in each direction. To do this, the outer loops for each reduction must be collapsed or \textit{coalesced} into a single loop with an upper bound given as a function of the upper bound of each of the outer loops. The induction variables corresponding to each of the outer loops must be recomputed inside coalesced loop body. For perfectly nested loops with constant upper bounds, such as any loop over the quadrature points or expansion modes for the hex, this transformation exist as an MLIR pass in the \texttt{affine} dialect, namely \texttt{affine-loop-coalescing}. For loops over the modes for a collapsed shape, where the upper bound is not constant, we have implemented our own transformation which implements loop coalescing for triangular and tetrahedral iteration spaces. A mathematical treatment of the method that inspired our implementation is given in~\cite{Clauss2017AutomaticLoops}. The loop coalescing is expected to occur before the \texttt{gpu.launch} operation replaces the loop over elements. Figure \ref{fig:loop coalescing} highlights this transformation applied to the triangular loop nest shown in Figure \ref{fig:bwd as loops}. As index computations are costly on GPU hardware, we generate index arrays that contain all the possible index values for the loops that are coalesced as part of our transformation. The induction variables for these loops are replaced by loads from the index arrays.   
This threading strategy does not require the expansion of the temporary data to the device warp size, and these can be placed in the device shared memory. The operands of the elemental operations can also be copied to the shared memory. To avoid bank conflicts and the serialization of accesses to the device shared memory, these shared memory allocations may need to be padded. 

In the next section, we will present the performance of the Helmholtz kernel generated for both threading strategies on the GPU, the performance of the kernel on the CPU as well as the compiler overhead introduced by the pipeline shown in Figure \ref{fig:loweringpipeline}.

\section{Performance evaluation}\label{results}
Here we present a preliminary investigation into the performance of our domain specific compiler. This falls into two categories: the overhead time, i.e. the time taken to generate, lower and JIT compile a kernel; and the runtime performance, i.e. the time taken for a kernel to execute once it is compiled. By far the more important is the latter quantity, since overheads are generally incurred only once at the start of a simulation, whereas runtime costs occur every time operator evaluations are required within the solver as part of a conjugate gradient or GMRES iteration. Nevertheless, measurement of overhead is important in e.g. adaptive simulations where performance-critical parameters may vary frequently at runtime.

\subsection{Overhead}
To quantify the compiler overhead, we measure the time taken for a kernel to be converted from NektarIR to LLVM, which we denote as the \textit{time to lower}, and the time it took for the kernel to be JIT compiled for the desired hardware target by the LLVM execution engine, which we refer to as the \textit{time to compile}. We consider these quantities for the Helmholtz operator on hexahedral and tetrahedral elements when targeting both CPU and GPU architectures for a range of polynomial orders with a fixed number of elements. In each case, the lowering and compilation times were computed as the mean of 1000 repetitions. On the CPU, this mean corresponds to the mean time to lower and time to compile across 128 MPI ranks on two AMD EPYC 9554 CPUs. The GPU target is a NVIDIA H100, with the time to lower and time to compile measured on the host. Compilation to GPU binaries was performed by version 12.8 of the NVCC CUDA compiler.

\begin{figure}
    \centering
    \includegraphics[width=\linewidth]{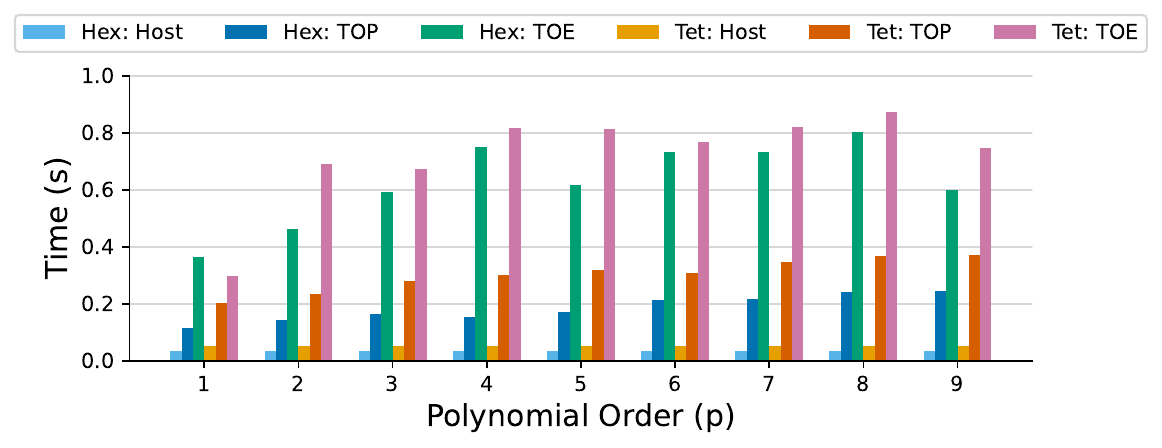}
    \caption{\textbf{Time to lower the Helmholtz operator on hexahedral and tetrahedral elements from NektarIR to the LLVM dialect for both host and device targets.} TOP and TOE refer to the two threading strategies, namely through the expansion modes and over the elements respectively.}
    \label{fig:time_to_lower}
    \Description[A bar graph of time to lower versus polynomial order]{For each polynomial order, the time to lower from NektarIR to LLVM IR for the Helmholtz operator on both hexahedral and tetrahedral elements is shown. The measurements include both the CPU and GPU times.}
\end{figure}

Figure~\ref{fig:time_to_lower} contains the mean time to lower and highlights the significant difference in the time to lower to a CPU target compared to a GPU target. This is expected as the compilation to GPU binaries is offloaded to the device compiler during the lowering stage and before the host component of the kernel is translated to LLVM IR, as highlighted in Figure~\ref{fig:loweringpipeline}. On the host side, the lowering pipeline is less than 0.1 seconds and on the GPU the lowering pipeline remains less than 0.9 seconds.  

\begin{figure}
    \centering
    \includegraphics[width=\linewidth]{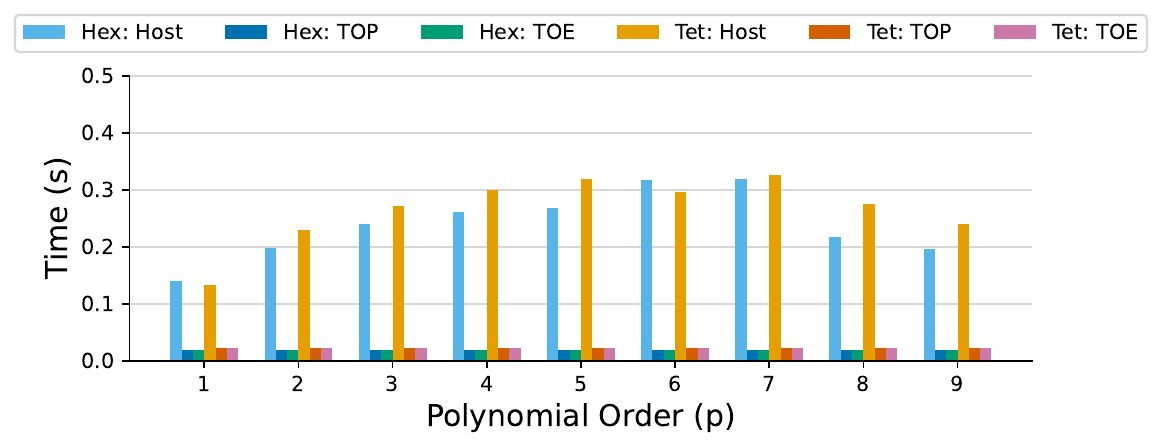}
    \caption{\textbf{Time to compile the Helmholtz operator on hexahedral and tetrahedral elements from NektarIR to the LLVM dialect for both host and device targets.} TOP and TOE refer to the two threading strategies, namely through the expansion modes and over the elements respectively.}
    \label{fig:time_to_compile}
    \Description[A bar graph of time to compile versus polynomial order]{For each polynomial order, the time to compile for the Helmholtz operator on both hexahedral and tetrahedral elements is shown. The measurements include both the CPU and GPU times.}
\end{figure}

Figure~\ref{fig:time_to_compile} shows the compilation stage for CPU kernels remains less than 0.35 seconds. As the majority of the GPU kernel has been compiled to the device target during the lowering stage, there is very little work to be done to compile host component of the kernel and the compilation time for GPU kernels remains below 0.05 seconds.  

These results show that NektarIR is able to generate and compile a Helmholtz kernel for a given target in less than 1 second. While this timescale may be considerable for computational workloads where the runtime is of a similar order, it is negligible in the context of CFD simulations which may require days to complete. Furthermore, these results highlight the flexibility of the JIT compilation approach and its suitability for adaptive simulations that require fast generation of new kernels as the polynomial order changes.

\subsection{Runtime Performance}
To quantify the runtime performance of the code-generated kernels, we measure the execution times for kernels across a range of polynomial orders and mesh sizes. The performance of these kernels has been extensively investigated in \cite{Moxey2020EfficientElements} and \cite{Xing2026Architecture-awareMeshes}, where we are interested in measuring the number of degrees of freedom that are processed per second, or the \emph{throughput}. For a given kernel, the throughput is given by
\begin{equation}
    \text{Throughput} = \frac{\text{\#(Degrees of Freedom)}}{\text{Execution Time}}
    \label{eqn: throughput}
\end{equation}
where the number of degrees of freedom is given by the (total number of input modes)$\times$(the number of elements). We compare the throughput to the corresponding kernel in the Nektar++ spectral/\textit{hp} element framework. 

\subsubsection{CPU}
\begin{figure}
    \centering
    \includegraphics[width=0.9\linewidth]{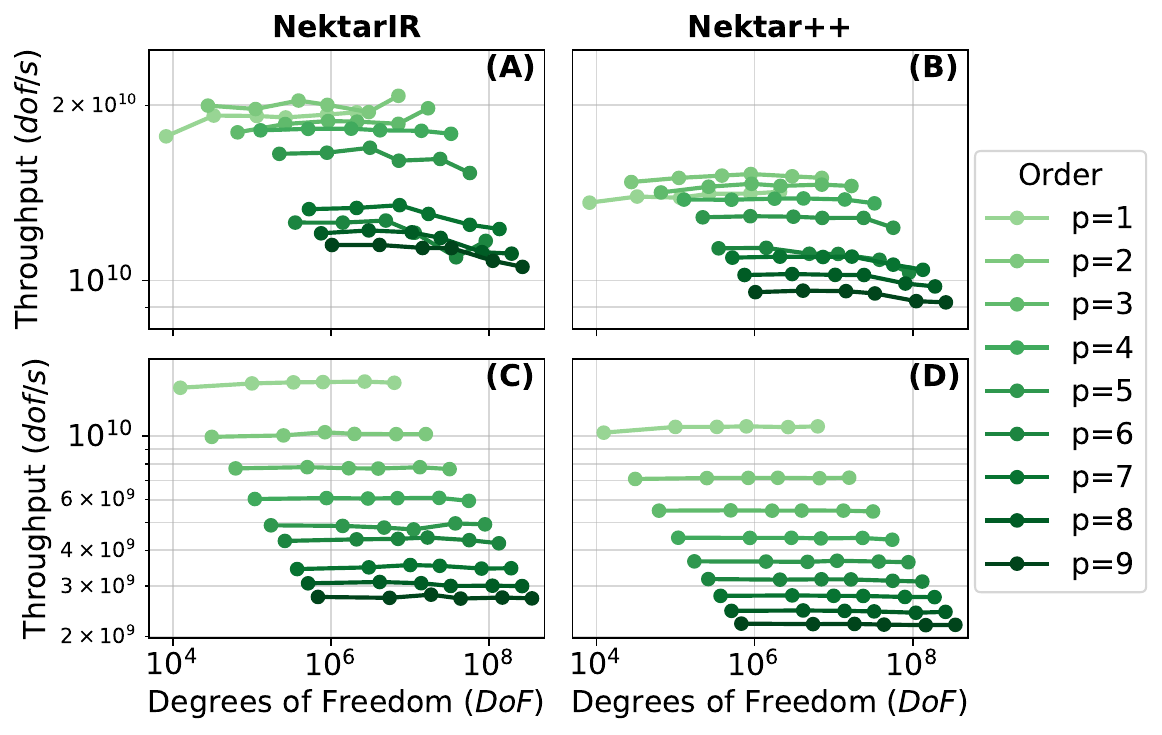}
    \caption{\textbf{Throughput comparison of the AVX512 Helmholtz kernel in NektarIR and Nektar++ on AMD EPYC 9554 CPU.} \textbf{(A)} and \textbf{(B)} show the throughput of the Helmholtz kernel on hexahedral elements while \textbf{(C)} and \textbf{(D)} correspond to the operator on tetrahedral elements. Each panel shows curves plotted on two logarithmic axes.}
    \label{fig:host_results}
    \Description[A collection of plots of the throughput verus degrees of freedom]{The figure contains 4 subplots, each showing the throughput versus degrees of freedom for the Helmholtz operator on CPU targets. Each subplot contains 9 curves corresponding to the polynomial orders from 1 to 9, inclusive.}
\end{figure}

The CPU runtime evaluation compares the throughput of the Helmholtz kernel generated for AVX512 instructions using NektarIR and the corresponding kernel in Nektar++. It is important to note that while the NektarIR kernels generate broadcast instructions for operands such as the basis and weights tensors, the Nektar++ kernels use a vector load instruction and store vector-width number of copies of these operands contiguously in memory instead. In both cases, the kernels are launched in parallel using MPI across 128 pinned ranks on two AMD EPYC 9554 CPUs. For these cases, NektarIR was compiled with \texttt{clang++} version 21.1.0. The Nektar++ results were obtained using the redesign branch\footnote{\url{https://gitlab.nektar.info/nektar/nektar}, feature/redesign, commit: 2e3e6bcb7d8559411e6a52824319f8577f54ce55 \label{nekver}} of Nektar++ compiled in Release mode with \texttt{gcc} version 11.4.0 with compile flags \texttt{-O3 -DNDEBUG -mfma -mavx512f}. The execution time computed as the average time for 1000 consecutive calls to the kernel. For the timing, we used the \texttt{chrono} library in C++. The efficiency of the Nektar++ kernels has been investigated in more detail in~\cite{Moxey2020EfficientElements}. Each MPI rank receives a segment of the total number of elements to act on and the NektarIR kernels are generated on each MPI rank individually. Figure~\ref{fig:host_results} shows the throughput comparison for both hexahedral and tetrahedral elements. The NektarIR pipeline used to generate the kernels is the same as the one shown in Figure \ref{fig:loweringpipeline}. While the throughput decreases as the polynomial order and computational work increases, panels \textbf{A} and \textbf{B} of Figure~\ref{fig:host_results} show the NektarIR kernel outperforms Nektar++ in all cases. For tetrahedral elements, panels \textbf{C} and \textbf{D} in Figure \ref{fig:host_results} shows a higher throughput for the NektarIR kernel at all polynomial orders.
 While the throughput advantage of NektarIR decreases as the polynomial order increases for both hexahedral and tetrahedral elements, the results shown in Figure~\ref{fig:host_results} highlight how the NektarIR pipeline is able to produce performant kernels for an AVX512 target using a modified vectorization strategy from Nektar++.

\subsubsection{GPU}
To match the GPU implementations in Nektar++, we consider kernels for both threading strategies outlined in Section~\ref{gpusection}, i.e. threading over elements and threading over expansion coefficients. The GPU kernels in Nektar++ have been developed as part of a recent redesign effort and are written as separate hand-written implementations for each threading strategy and element type, with performance critical parameters explicitly compiled via C++ template arguments. All benchmarks were run on an NVIDIA H100 GPU. The GPU kernels were compiled with version 12.8 of the NVCC CUDA compiler with \texttt{sm\_90} as the target architecture with the \texttt{-O3} and \texttt{-DNDEBUG} flags. The Nektar++ redesign branch\footref{nekver} was compiled with \texttt{gcc} version 11.4.0, while NektarIR was compiled using \texttt{clang++} version 21.1.0. To isolate the time spent by the device executing the instructions generated for both kernels, we used the Nvidia Nsight Compute profiler. In all cases, the kernels were allowed to run for 200 iterations to warm up the device before the profiler collected the \texttt{gpu\_\_time\_\_duration.sum} execution time metric. The throughput was then computed using equation~\eqref{eqn: throughput}.

\begin{figure}
    \centering
    \includegraphics[width=0.95\linewidth]{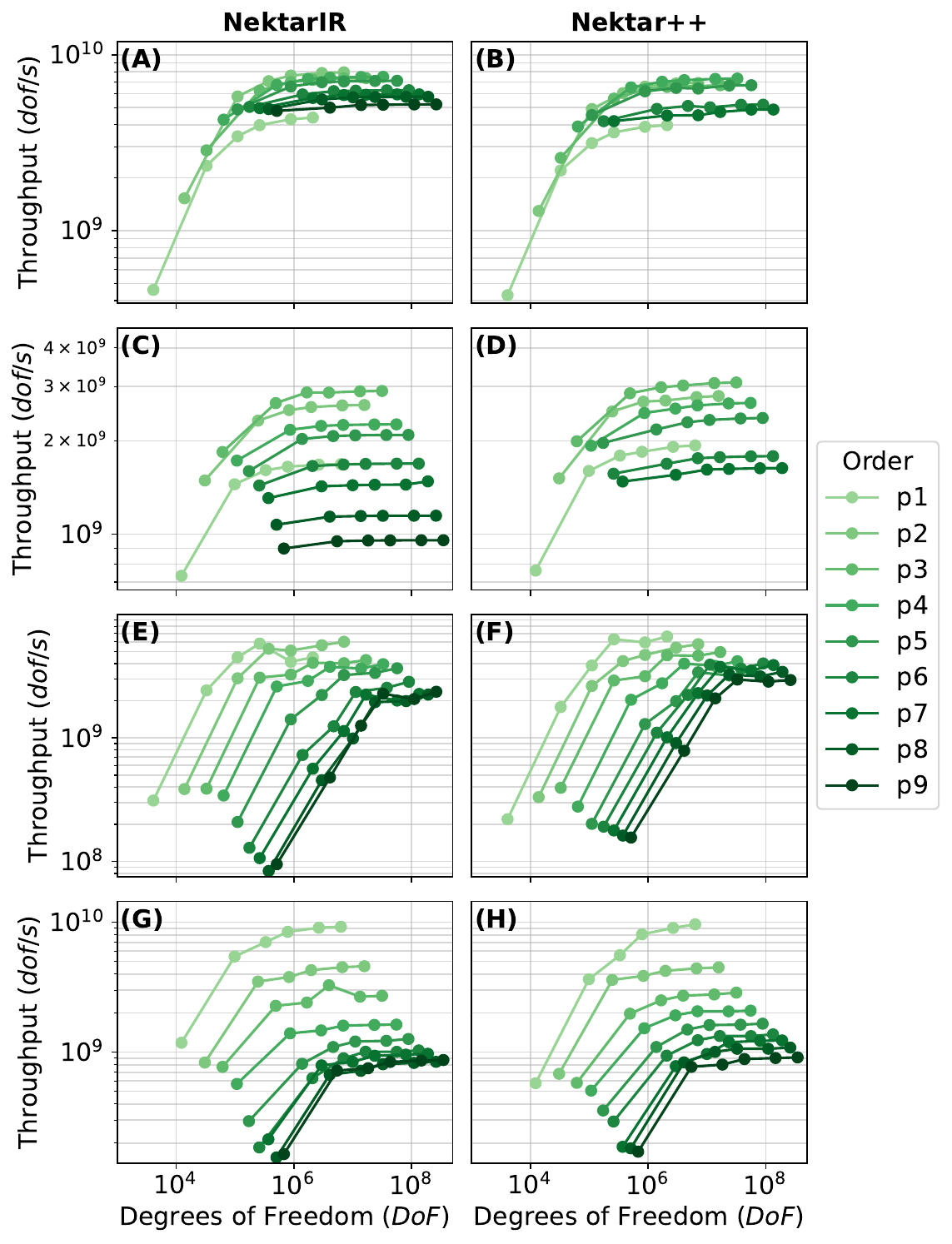}
    \caption{\textbf{Throughput comparison of the Helmholtz kernel in NektarIR and Nektar++ on a NVIDIA H100 GPU.} \textbf{(A)-(D)} correspond to the threading through expansion mode method on \textbf{(A)-(B)} hexahedral and \textbf{(C)-(D)} tetrahedral elements. \textbf{(E)-(H)} correspond to the threading through elements method on \textbf{(E)-(F)} hexahedral and \textbf{(G)-(H)} tetrahedral elements. Each panel shows curves plotted on two logarithmic axes.}
    \Description[A collection of plots of the throughput verus degrees of freedom]{The figure contains 8 subplots, each showing the throughput versus degrees of freedom for the Helmholtz operator on GPU targets, for both hexahedral and tetrahedral elements and both of the above mentioned threading strategies. Each subplot contains 9 curves corresponding to the polynomial orders from 1 to 9, inclusive.}
    \label{fig:GPURESULTS}
\end{figure}

Figure~\ref{fig:GPURESULTS} shows a comparison of the throughput obtained by the Helmholtz operator in NektarIR and Nektar++ with both the threading strategies discussed in Section~\ref{gpusection} on both hexahedral and tetrahedral elements. For the threading through expansion mode and threading over elements approaches, the throughput for the Helmholtz kernel on hexahedral elements is comparable for the small to medium number of degrees of freedom. However, for large problem sizes with degrees of freedom of the order near $10^{8}$, the discrepancy between NektarIR and Nektar++ grows. The NSight Compute profile for these kernels suggest that the performance deficit occurs due to an increased register pressure for the NektarIR kernels, however the source of this in the IR remains to be found and is a target of further study. The results for the threading over elements strategy on tetrahedral elements shows a similar trend. On tetrahedral elements, the threading through expansion mode method shows a considerably worse throughput in the kernels generated using NektarIR compared to those from Nektar++. Addressing this performance discrepancy remains a target for future study.  

\section{Conclusion}\label{conclusion}
In this work we have presented NektarIR, a domain-specific compiler for code-generation of high-order finite element kernels for the spectral/\textit{hp} element method. These kernels are crucial for the performance of any spectral/\textit{hp} element solver in computational fluid dynamics. By creating our own MLIR dialect with a high level abstraction of the elemental operations, NektarIR leverages domain-specific IR rewriting to generate hardware specific kernels from a single representation. After lowering the IR through preexisting MLIR dialects, the NektarIR kernels are translated to LLVM IR and leverage the LLVM compiler infrastructure's extensive hardware support for extensibility. By measuring the compiler overhead, we showed that NektarIR can generate kernels for GPU targets in less than 1 second, a time-scale which is negligible compared to the runtime duration of CFD simulations. We illustrated the end-to-end compilation pipeline of our kernels from a C++ front-end to both CPU and GPU architectures and compared the runtime performance to kernels from the Nektar++ spectral/\textit{hp} element framework for high-fidelity fluid dynamics simulations. This comparison shown performance benefits to NektarIR on the CPU while the GPU performance requires further optimization to match the hand-optimized kernels in Nektar++.

While this work has presented the method used in NektarIR and the framework, a considerable amount of optimization opportunities remain possible, both from the spectral/\textit{hp} element method and via IR transformations using MLIR. Although the current goal of NektarIR is to facilitate the generation of the elemental operations for the use in spectral/\textit{hp} element solvers, extending the abstraction in NektarIR to facilitate the generation of a full CFD solver remains possible as part of future work. Other high-order CFD frameworks such as Neko \cite{Jansson2024Neko:Dynamics} and MFEM \cite{Andrej2024High-performanceMFEM} utilize different abstractions of the spectral/\textit{hp} element method to achieve high-performance CFD simulations on both CPU and GPU hardware and investigating how their abstractions can be represented using NektarIR is a possible avenue of future study. Extending the elemental operations to support elemental expansions using other basis types, such as Lagrange polynomials, would further extend the type of fluid dynamics that can be simulated using a NektarIR based solver. 

While the elemental operations represented in NektarIR are specific to the spectral/\textit{hp} element method, they are mathematically similar to batched dense-dense or dense-sparse matrix multiplications. There is considerable support for operations of this kind in MLIR already via for instance the \texttt{linalg} and \texttt{sparse} dialects \cite{Bik2022CompilerMLIR} and facilitating the conversion from our dialect, \texttt{nir}, to these may enable the utilization of more optimizing transformations during the lowering pipeline for our IR.

\bibliographystyle{ACM-Reference-Format}
\bibliography{references}

@book{sherwinbook,
    title = {{Spectral/hp Element Methods for Computational Fluid Dynamics}},
    year = {2005},
    author = {Karniadakis, George and Sherwin, Spencer},
    edition = {2nd},
    month = {6},
    publisher = {Oxford University Press},
    address={Oxford, United Kingdom},
    url = {https://doi.org/10.1093/acprof:oso/9780198528692.001.0001},
    isbn = {9780198528692},
    doi = {10.1093/acprof:oso/9780198528692.001.0001}
}

@book{LLVMBook,
    title = {{The architecture of open source applications : elegance, evolution, and a few fearless hacks}},
    year = {2011},
    author = {Brown, Amy. and Wilson, Greg.},
    pages = {415},
    publisher = {[CreativeCommons]},
    address={CA, USA},
    url = {https://aosabook.org/en/v1/llvm.html},
    isbn = {9781257638017}
}

@misc{Xing2026Architecture-awareMeshes,
    title={Architecture-aware $h$-to-$p$ optimisation: spectral/$hp$ element operators for mixed-element meshes}, 
      author={Jacques Y. Xing and Boyang Xia and Diego Renner and Chris D. Cantwell and David Moxey and Robert M. Kirby and Spencer J. Sherwin},
      year={2026},
      eprint={2604.04644},
      archivePrefix={arXiv},
      primaryClass={math.NA},
      url={https://arxiv.org/abs/2604.04644}, 
}

@inproceedings{Clauss2017AutomaticLoops,
    title = {{Automatic Collapsing of Non-Rectangular Loops}},
    year = {2017},
    booktitle = {Proceedings - 2017 IEEE 31st International Parallel and Distributed Processing Symposium, IPDPS 2017},
    ADDRESS = {Orlando, United States},
    PUBLISHER = {{IEEE International}},
    PAGES = {778 - 787},
    author = {Clauss, Philippe and Altintas, Ervin and Kuhn, Matthieu},
    doi = {10.1109/IPDPS.2017.34}
}

@online{ClangDevelopersClang:LLVM,
    title = {{Clang: a C language family frontend for LLVM}},
    key = {{Clang}},
    url = {https://clang.llvm.org/},
    lastaccessed= "1 Jun 2026"
}

@article{Bik2022CompilerMLIR,
    title = {{Compiler Support for Sparse Tensor Computations in MLIR}},
    year = {2022},
    journal = {ACM Transactions on Architecture and Code Optimization},
    author = {Bik, Aart and Koanantakool, Penporn and Shpeisman, Tatiana and Vasilache, Nicolas and Zheng, Bixia and Kjolstad, Fredrik},
    number = {4},
    month = {9},
    pages = {1--25},
    volume = {19},
    publisher = {Association for Computing Machinery},
    doi = {10.1145/3544559},
    issn = {15443973},
    arxivId = {2202.04305}
}

@article{Baratta2025DOLFINx:Environment,
    title = {{DOLFINx: The next generation FEniCS problem solving environment}},
    year = {2025},
    journal = {10.5281/zenodo.10447665.},
    author = {Baratta, Igor A and Dean, Joseph P and Dokken, Jørgen S and Hale, Jack S and Richardson, Chris N and Rognes, Marie E and Scroggs, Matthew W and Sime, Nathan and Wells, Garth N},
    month = {12},
    doi = {10.5281/zenodo.18101307}
}

@article{Gysi2021Domain-SpecificSimulation,
    title = {{Domain-Specific Multi-Level IR Rewriting for GPU: The Open Earth Compiler for GPU-Accelerated Climate Simulation}},
    year = {2021},
    journal = {ACM Transactions on Architecture and Code Optimization},
    author = {Gysi, Tobias and M{\"{u}}ller, Christoph and Zinenko, Oleksandr and Herhut, Stephan and Davis, Eddie and Wicky, Tobias and Fuhrer, Oliver and Hoefler, Torsten and Grosser, Tobias},
    number = {4},
    month = {12},
    volume = {18},
    pages={1--23},
    publisher = {Association for Computing Machinery},
    doi = {10.1145/3469030},
    issn = {15443973}
}

@article{Kolev2021EfficientMethods,
    title = {{Efficient exascale discretizations: High-order finite element methods}},
    year = {2021},
    journal = {International Journal of High Performance Computing Applications},
    author = {Kolev, Tzanio and Fischer, Paul and Min, Misun and Dongarra, Jack and Brown, Jed and Dobrev, Veselin and Warburton, Tim and Tomov, Stanimire and Shephard, Mark S. and Abdelfattah, Ahmad and Barra, Valeria and Beams, Natalie and Camier, Jean Sylvain and Chalmers, Noel and Dudouit, Yohann and Karakus, Ali and Karlin, Ian and Kerkemeier, Stefan and Lan, Yu Hsiang and Medina, David and Merzari, Elia and Obabko, Aleksandr and Pazner, Will and Rathnayake, Thilina and Smith, Cameron W. and Spies, Lukas and Swirydowicz, Kasia and Thompson, Jeremy and Tomboulides, Ananias and Tomov, Vladimir},
    number = {6},
    month = {11},
    pages = {527--552},
    volume = {35},
    publisher = {SAGE Publications Inc.},
    doi = {10.1177/10943420211020803},
    issn = {17412846},
    arxivId = {2109.04996}
}

@article{Moxey2020EfficientElements,
    title = {{Efficient matrix-free high-order finite element evaluation for simplicial elements}},
    year = {2020},
    journal = {SIAM Journal on Scientific Computing},
    author = {Moxey, David and Amici, Roman and Kirby, Mike},
    number = {3},
    pages = {C97-C123},
    volume = {42},
    publisher = {Society for Industrial and Applied Mathematics Publications},
    doi = {10.1137/19M1246523},
    issn = {10957197}
}

@article{Eichstadt2023EfficientDimensions,
    title = {{Efficient vectorised kernels for unstructured high-order finite element fluid solvers on GPU architectures in two dimensions}},
    year = {2023},
    journal = {Computer Physics Communications},
    author = {Eichst{\"{a}}dt, Jan and Peir{\'{o}}, Joaquim and Moxey, David},
    month = {3},
    volume = {284},
    publisher = {Elsevier B.V.},
    doi = {10.1016/j.cpc.2022.108624},
    issn = {00104655},
    pages= {108624}
}

@article{Rathgeber2016Firedrake:Abstractions,
    title = {{Firedrake: Automating the finite element method by composing abstractions}},
    year = {2016},
    pages={1--27},
    journal = {ACM Transactions on Mathematical Software},
    author = {Rathgeber, Florian and Ham, David A. and Mitchell, Lawrence and Lange, Michael and Luporini, Fabio and McRae, Andrew T.T. and Bercea, Gheorghe Teodor and Markall, Graham R. and Kelly, Paul H.J.},
    number = {3},
    volume = {43},
    doi = {10.1145/2998441},
    issn = {15577295}
}

@article{Kirilov2024High-OrderMeshes,
    title = {High-order curvilinear mesh generation from third-party meshes},
    journal = {Computer-Aided Design},
    volume = {191},
    pages = {103962},
    year = {2026},
    issn = {0010-4485},
    doi = {https://doi.org/10.1016/j.cad.2025.103962},
    url = {https://www.sciencedirect.com/science/article/pii/S001044852500123X},
    author = {Kaloyan S. Kirilov and Jingtian Zhou and Joaquim Peiró and David Moxey}
}

@article{Karniadakis1991High-orderEquations,
    title = {{High-order splitting methods for the incompressible Navier-Stokes equations}},
    year = {1991},
    pages = {414--443},
    journal = {Journal of Computational Physics},
    author = {Karniadakis, George Em and Israeli, Moshe and Orszag, Steven A.},
    number = {2},
    volume = {97},
    doi = {10.1016/0021-9991(91)90007-8},
    issn = {10902716}
}

@article{Andrej2024High-performanceMFEM,
    title = {{High-performance finite elements with MFEM}},
    year = {2024},
    journal = {International Journal of High Performance Computing Applications},
    author = {Andrej, Julian and Atallah, Nabil and B{\"{a}}cker, Jan Phillip and Camier, Jean Sylvain and Copeland, Dylan and Dobrev, Veselin and Dudouit, Yohann and Duswald, Tobias and Keith, Brendan and Kim, Dohyun and Kolev, Tzanio and Lazarov, Boyan and Mittal, Ketan and Pazner, Will and Petrides, Socratis and Shiraiwa, Syun’ichi and Stowell, Mark and Tomov, Vladimir},
    number = {5},
    pages = {447--467},
    volume = {38},
    doi = {10.1177/10943420241261981},
    issn = {17412846}
}

@article{Bezanson2017Julia:Computing,
    title = {{Julia: A fresh approach to numerical computing}},
    year = {2017},
    journal = {SIAM Review},
    author = {Bezanson, Jeff and Edelman, Alan and Karpinski, Stefan and Shah, Viral B.},
    number = {1},
    pages = {65--98},
    volume = {59},
    doi = {10.1137/141000671},
    issn = {00361445}
}

@inproceedings{Lattner2004LLVM:Transformation,
    title = {{LLVM: A Compilation Framework for Lifelong Program Analysis {\&} Transformation}},
    pages={75},
    year = {2004},
    booktitle = {CGO '04: Proceedings of the international symposium on Code generation and optimization: feedback-directed and runtime optimization},
    author = {Lattner, Chris and Adve, Vikram},
    month = {3},
    publisher = {IEEE Computer Society},
    url = {http://llvm.cs.uiuc.edu/},
    address = {Palo Alto}
}

@article{Lattner2020MLIR:Law,
    title = {{MLIR: A Compiler Infrastructure for the End of Moore's Law}},
    year = {2020},
    pages ={1--21},
    journal = {CoRR},
    author = {Lattner, Chris and Pienaar, Jacques A and Amini, Mehdi and Bondhugula, Uday and Riddle, River and Cohen, Albert and Shpeisman, Tatiana and Davis, Andy and Vasilache, Nicolas and Zinenko, Oleksandr},
    volume = {abs/2002.11054},
    url = {https://arxiv.org/abs/2002.11054}
}

@article{Jansson2024Neko:Dynamics,
    title = {{Neko: A modern, portable, and scalable framework for high-fidelity computational fluid dynamics}},
    year = {2024},
    pages={106243},
    journal = {Computers and Fluids},
    author = {Jansson, Niclas and Karp, Martin and Podobas, Artur and Markidis, Stefano and Schlatter, Philipp},
    volume = {275},
    doi = {10.1016/j.compfluid.2024.106243},
    issn = {00457930}
}

@article{Cantwell2015Nektar++:Framework,
    title = {{Nektar++: An open-source spectral/hp element framework}},
    year = {2015},
    journal = {Computer Physics Communications},
    author = {Cantwell, C. D. and Moxey, D. and Comerford, A. and Bolis, A. and Rocco, G. and Mengaldo, G. and De Grazia, D. and Yakovlev, S. and Lombard, J. E. and Ekelschot, D. and Jordi, B. and Xu, H. and Mohamied, Y. and Eskilsson, C. and Nelson, B. and Vos, P. and Biotto, C. and Kirby, R. M. and Sherwin, S. J.},
    month = {7},
    pages = {205--219},
    volume = {192},
    publisher = {Elsevier B.V.},
    doi = {10.1016/j.cpc.2015.02.008},
    issn = {00104655}
}

@article{Moxey2020Nektar++:,
    title = {{Nektar++: Enhancing the capability and application of high-fidelity spectral/hp element methods}},
    year = {2020},
    journal = {Computer Physics Communications},
    author = {Moxey, D and Cantwell, C D and Bao, Y},
    pages = {107110},
    volume = {249},
    url = {http://dx.doi.org/10.17632/9drxd9d8nx.1CodeOceanCapsule:https://doi.org/10.24433/CO.9865757.v1 http://creativecommons.org/licenses/by/4.0/},
    doi = {10.17632/9drxd9d8nx.1}
}

@article{Orszag1980SpectralGeometries,
    title = {{Spectral methods for problems in complex geometries}},
    year = {1980},
    pages={70--92},
    journal = {Journal of Computational Physics},
    author = {Orszag, Steven A.},
    number = {1},
    volume = {37},
    doi = {10.1016/0021-9991(80)90005-4},
    issn = {00219991}
}

@article{Bastian2021TheDevelopments,
    title = {{The DUNE framework: Basic concepts and recent developments}},
    year = {2021},
    journal = {Computers and Mathematics with Applications},
    author = {Bastian, Peter and Blatt, Markus and Dedner, Andreas and Dreier, Nils Arne and Engwer, Christian and Fritze, René and Gr{\"{a}}ser, Carsten and Gr{\"{u}}ninger, Christoph and Kempf, Dominic and Kl{\"{o}}fkorn, Robert and Ohlberger, Mario and Sander, Oliver},
    pages = {75--112},
    volume = {81},
    doi = {10.1016/j.camwa.2020.06.007},
    issn = {08981221}
}

@article{Liu2022TinyIREE:Deployment,
    title = {{TinyIREE: An ML Execution Environment for Embedded Systems from Compilation to Deployment}},
    year = {2022},
    pages={9--16},
    journal = {IEEE Micro},
    author = {Liu, Hsin I.Cindy and Brehler, Marius and Ravishankar, Mahesh and Vasilache, Nicolas and Vanik, Ben and Laurenzo, Stella},
    number = {5},
    volume = {42},
    doi = {10.1109/MM.2022.3178068},
    issn = {19374143}
}

@Article{2025:dealII,
  author  = {Daniel Arndt and Wolfgang Bangerth and Maximilian Bergbauer and Bruno Blais and Marc Fehling and Rene Gassm\"{o}ller
             and Timo Heister and Luca Heltai and Martin Kronbichler and Matthias Maier and Peter Munch and Sam Scheuerman
             and Bruno Turcksin and Siarhei Uzunbajakau and David Wells and Micha{\l} Wichrowski},
  title   = {The deal.II library, Version 9.7},
  journal = {Journal of Numerical Mathematics},
  year    = 2025,
  volume  = 33,
  number  = 4,
  pages   = {403--415},
  doi     = {10.1515/jnma-2025-0115}
}

@article{2022:NekRS,
  title = {NekRS, a GPU-accelerated spectral element Navier–Stokes solver},
    journal = {Parallel Computing},
    volume = {114},
    pages = {102982},
    year = {2022},
    issn = {0167-8191},
    doi = {https://doi.org/10.1016/j.parco.2022.102982},
    url = {https://www.sciencedirect.com/science/article/pii/S0167819122000710},
    author = {Paul Fischer and Stefan Kerkemeier and Misun Min and Yu-Hsiang Lan and Malachi Phillips and Thilina Rathnayake and Elia Merzari and Ananias Tomboulides and Ali Karakus and Noel Chalmers and Tim Warburton}
}

@article{hp-adaptivepaper1,
author = {Mossier, Pascal and Appel, Daniel and Beck, Andrea D. and Munz, Claus-Dieter},
title = {An Efficient hp-Adaptive Strategy for a Level-Set Ghost-Fluid Method},
year = {2023},
issue_date = {Nov 2023},
publisher = {Plenum Press},
address = {USA},
volume = {97},
number = {2},
issn = {0885-7474},
url = {https://doi.org/10.1007/s10915-023-02363-7},
doi = {10.1007/s10915-023-02363-7},
journal = {J. Sci. Comput.},
month = oct,
numpages = {41}
}

@article{hpadaptivepaper2, 
author = {Fehling, Marc and Bangerth, Wolfgang},
title = {Algorithms for Parallel Generic hp-Adaptive Finite Element Software},
year = {2023},
issue_date = {September 2023},
publisher = {Association for Computing Machinery},
address = {New York, NY, USA},
volume = {49},
number = {3},
issn = {0098-3500},
url = {https://doi.org/10.1145/3603372},
doi = {10.1145/3603372},
journal = {ACM Trans. Math. Softw.},
month = sep,
articleno = {25},
numpages = {26}
}

@article{roofline,
author = {Williams, Samuel and Waterman, Andrew and Patterson, David},
title = {Roofline: an insightful visual performance model for multicore architectures},
year = {2009},
issue_date = {April 2009},
publisher = {Association for Computing Machinery},
address = {New York, NY, USA},
volume = {52},
number = {4},
issn = {0001-0782},
url = {https://doi.org/10.1145/1498765.1498785},
doi = {10.1145/1498765.1498785},
journal = {Commun. ACM},
month = apr,
pages = {65–76},
numpages = {12}
}

@article{reviewSIAM,
author = {Mengaldo, Gianmarco and Moxey, David and Turner, Michael and Moura, Rodrigo Costa and Jassim, Ayad and Taylor, Mark and Peir\'{o}, Joaquim and Sherwin, Spencer},
title = {Industry-Relevant Implicit Large-Eddy Simulation of a High-Performance Road Car via Spectral/hp Element Methods},
journal = {SIAM Review},
volume = {63},
number = {4},
pages = {723-755},
year = {2021},
doi = {10.1137/20M1345359},

URL = { 
    
        https://doi.org/10.1137/20M1345359
    
    

},
eprint = { 
    
        https://doi.org/10.1137/20M1345359
    
    

}
}

@online{stablehlo,
    author = {OpenXLA Community},
    title = {StableHLO Specification},
    year = {2023},
    howpublished = {\url{https://openxla.org/stablehlo/spec}},
    note = {Accessed: 1 Jun 2026}
}

@online{LLVM_Torch-MLIR,
author = {{LLVM}},
license = {["Apache-2.0 with LLVM Exceptions", "BSD"]},
title = {{Torch-MLIR}},
howpublished = {\url{https://github.com/llvm/torch-mlir}},
note = {Accessed: 1 Jun 2026}
}

@book{klabnik2026rust,
  title={The Rust Programming Language, 3rd Edition},
  author={Klabnik, S. and Nichols, C. and Krycho, C.},
  isbn={9781718504448},
  url={https://books.google.co.uk/books?id=Nm9REQAAQBAJ},
  year={2026},
  publisher={No Starch Press}
}
\end{document}